\RequirePackage{etex}
\documentclass[AMA,STIX1COL]{WileyNJD-v2} 
\usepackage{amsmath}
\usepackage{amsxtra}
\usepackage{amstext}
\usepackage{amssymb}
\usepackage{latexsym}
\usepackage{dsfont} 

\makeatletter
\newcommand\makebig[2]{%
  \@xp\newcommand\@xp*\csname#1\endcsname{\bBigg@{#2}}%
  \@xp\newcommand\@xp*\csname#1l\endcsname{\@xp\mathopen\csname#1\endcsname}%
  \@xp\newcommand\@xp*\csname#1r\endcsname{\@xp\mathclose\csname#1\endcsname}%
}
\makeatother

\makebig{biggg} {1.0}
\makebig{Biggg} {1.5}
\makebig{bigggg}{2.0}
\makebig{Bigggg}{0.75}
\makebig{gif}{1.4}
\makebig{giff}{1.25}
\makebig{smallBracket}{0.5}

\articletype{Research Article}%
\begin{document}

\title{Joint Hybrid Precoding and Multi-IRS Optimization for mmWave MU-MISO Communication Network}

\author[1]{Fardad Rahkheir}
\author[1]{Soroush Akhlaghi*}
\authormark{Joint Hybrid Precoding and Multi-IRS Optimization for mmWave MU-MISO Communication Network}

\address{\orgdiv{Department of Electrical Engineering}, \orgname{Shahed University}, \orgaddress{\state{Tehran}, \country{Iran}}}
\corres{*Soroush Akhlaghi, Department of Electrical Engineering, Shahed University, Tehran, Iran. \email{akhlaghi@shahed.ac.ir}}
\abstract[Abstract]{\textcolor{black}{This paper attempts to jointly optimize the hybrid precoding (HP) and intelligent reflecting surfaces (IRS) beamforming matrices in a multi-IRS-aided mmWave communication network, utilizing the Alamouti scheme at the base station (BS). Considering the overall signal-to-noise ratio (SNR) as the objective function, the underlying problem is cast as an optimization problem, which is shown to be non-convex in general. To tackle the problem, noting that the unknown matrices contribute multiplicatively to the objective function, they are reformulated into two new matrices with rank constraints. Then, using the so-called inner approximation (IA) technique in conjunction with majorization-minimization (MM) approaches, these new matrices are solved iteratively. From one of these matrices, the IRS beamforming matrices can be effectively extracted. Meanwhile, HP precoding matrices can be solved separately through a new optimization problem aimed at minimizing the Euclidean distance between the fully digital (FD) precoder and HP analog/digital precoders. This is achieved through the use of a modified block coordinate descent (MBCD) algorithm. Simulation results demonstrate that the proposed algorithm outperforms various benchmark schemes in terms of achieving a higher achievable rate.}}

\maketitle

\section{Introduction}\label{sec1}
\;\;\;With the explosive growth in mobile traffic demand, the gap between capacity needs and spectrum availability is becoming more critical. Millimeter-wave (mmWave) communication has recently attracted much attention from researchers and industry professionals due to its successful use in limited-space communication and the military \cite{ref1, ref2}. mmWave communications can achieve transmission speeds of up to 1 Gbps at 60 GHz, making it a promising solution for the emerging cellular and wireless generations \cite{ref2}. Nevertheless, there are several challenges associated with its implementation. Firstly, mmWave communications suffer from severe propagation loss due to their high carrier frequency and inherently directional properties \cite{ref1}. Beamforming (BF) can be adopted as an essential technique to mitigate mismatches between the transmit steering vector and the receive beamformer. Secondly, mmWave communications are sensitive to blockage by obstacles such as humans and trees due to their weak diffraction ability \cite{ref1}.\\ 
To tackle the former issue, fully digital (FD) precoding can be employed, while the prohibitive cost and power consumption of the hardware components in mmWave bands remain subjects of debate. The more profitable choice to reduce the number of power-hungry radio frequency (RF) chains at the base station (BS) would be to employ hybrid analog-digital precoding (HP) \cite{ref3}. To address the latter issue, we can benefit from relay-aided communication networks, which may not be cost-effective. Recent developments in electromagnetic (EM) metasurfaces have made intelligent reflecting surfaces (IRSs) a promising alternative to relay-based systems in terms of both cost and energy efficiency \cite{ref4}. Notably, the reflecting elements, which are composed of inexpensive programmable passive components such as phase shifters and positive-intrinsic-negative (PIN) diodes, enable the IRSs to adjust the reflection behavior of incoming electromagnetic (EM) signals \cite{ref4}. IRSs smartly adjust the phase shifts (PSs) of the passive elements; thus, the reflected signals can be combined coherently at the desired receiver to improve signal power or destructively at non-intended receivers to suppress interference \cite{ref5}. In other words, IRSs can be deployed to create a virtual line-of-sight (LoS) link between the users and their serving BS, bypassing any obstacles between them, especially for users in a service dead zone. This is particularly useful for coverage extension in mmWave communications, which are highly vulnerable to blockage \cite{ref6}.\\ 
Due to the benefits that IRS and HP provide, there have been some efforts to design single IRS-aided multi-user multi-input single-output (MU-MISO) mmWave communication systems \cite{ref7, ref8, ref9, ref10, ref11}. For instance, the authors of \cite{ref7} aim to minimize the mean-squared error (MSE) between the received symbols and the transmitted symbols by alternatively optimizing the analog-digital precoders at the BS and the phase shifts at the IRS, where the gradient-projection (GP) method is used to tackle the non-convex element-wise constant-modulus constraints for the analog precoding and the PSs. To date, there has been limited research on double/multi-IRS-aided mmWave communication systems \cite{ref12, ref13, ref14, ref15}, since increasing the number of IRSs may significantly impact the complexity of the optimization problem. As an example, the authors of \cite{ref12} aim to maximize the weighted sum rate (WSR) of a double IRS-aided hybrid beamforming architecture for the MU-MISO mmWave communication system, where they alternatively optimize the analog beamforming matrix at the BS and passive beamforming matrices at IRS 1 and IRS 2.\\ \\
On the other hand, Orthogonal Space-Time Block Codes (OSTBCs) are a powerful technique for improving the reliability and performance of wireless communication systems. OSTBCs based on orthogonal designs provide full diversity and maximum likelihood decoding with linear decoding complexity \cite{ref16}. This significant property allows for simplified receiver processing while maintaining robust performance in fading environments. Additionally, OSTBCs are highly adaptable to Multi-Input-Multi-Output (MIMO) systems, making them appropriate for different deployment scenarios. The use of orthogonal design ensures that all the symbols transmitted from different antennas are orthogonal to each other, which leads to effective interference mitigation and full diversity gain \cite{ref17}. Recently, the authors of \cite{ref18} and \cite{ref19} both considered a single IRS-aided communication network with the help of the Alamouti scheme at the BS, while the authors of \cite{ref19} do not consider a hybrid precoding structure at the transmitter.\\ \\
In this paper, we consider a hybrid analog/digital precoder and the Alamouti scheme in a MU-MISO mmWave communication system, aided by a double IRS implemented through programmable PSTs, which can be easily extended to a Multi-IRS-aided network. Specifically, we aim to jointly optimize the hybrid precoder at the BS after using the Alamouti encoder and the PSs at both IRSs such that the signal-to-noise ratio (SNR) is maximized. The resulting non-convex optimization problem is solved in two sub-problems. First, we jointly optimize the PSs at both IRSs and the full-digital precoder via the so-called inner approximation (IA) and penalty methods. Then, analog and hybrid precoders are optimized using the modified block coordinate descent (MBCD) algorithm proposed in \cite{ref20}. Numerical results reveal the desirable performance gain for the proposed joint hybrid precoding design compared to other conventional approaches. Through simulations, it is shown that the proposed Alamouti scheme outperforms the benchmark scheme in \cite{ref7} in terms of the achievable rate for the SNR region of interest, demonstrating the effectiveness of the proposed approach.\\ \\

\textit{Notations}: $y$, $\mathbf{y}$ and $\mathbf{Y}$ denote scalar, vector and matrix, respectively;  Transpose, conjugate transpose and pseudo inverse operators are represented by $\mathbf{Y}^T$, $\mathbf{Y}^H$ and $\mathbf{Y}^\dagger$, respectively; $\|\mathbf{Y}\|_F$, $\|\mathbf{Y}\|_2$, $\|\mathbf{Y}\|_*$ denote Frobenius, Spectral and Nuclear norms, respectively; $\text{diag}(\mathbf{y})$ operator diagonalizes $\mathbf{y}$; $\text{diagC}(\mathbf{Y})$ operator extracts the diagonal elements of $\mathbf{Y}$ into a column vector; $\text{Tr}(\mathbf{Y})$ and $\text{rank}(\mathbf{Y})$ denote trace and rank operator, respectively; Expectation of a random variable is noted by $\mathds{E}[\cdot]$; $|\cdot|$ returns the absolute value of a complex number; $\mathbf{I}_m \in \mathbb{C}^{m \times m}$ is the identity matrix; $\mathbf{1}_{m \times n}$ and $\mathbf{0}_{m \times n}$ are $m \times n$ matrices with all elements equal to 1 and 0, respectively; $\mathbf{X} \succeq 0$ stands for a positive semi-definite $\mathbf{X}$; $\mathbf{X}_{(a:b,m:n)}$ is a sub-matrix of $\mathbf{X}$ with rows $a$ to $b$ and columns $m$ to $n$; the symbol $\odot$ represents the Hadamard (element-wise) product; $\lambda_i$ represents the $i^{th}$ eigenvalue of a matrix. \\ \\

\section{System Model}
\begin{figure}
	\begin{center}
		\includegraphics[scale=0.14]{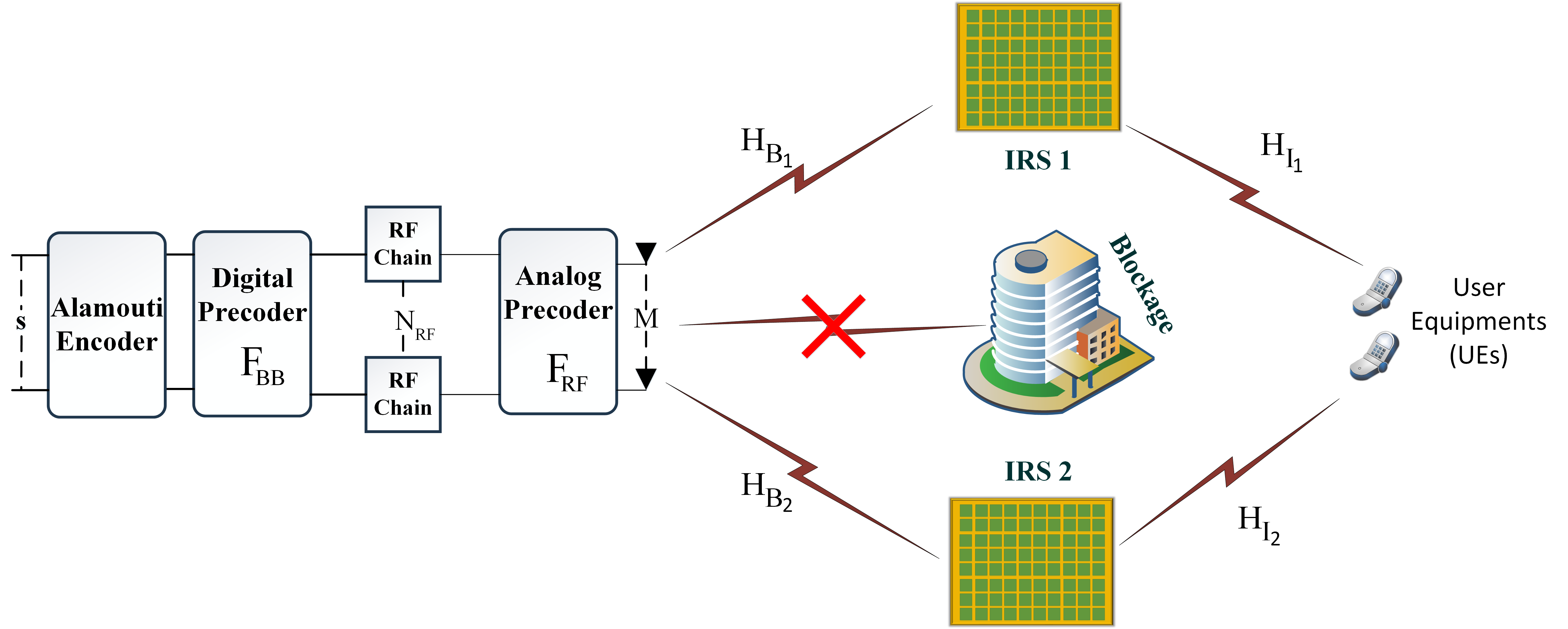}\vspace{-1mm}
		\caption{\;\;Double IRS-aided MU-MISO mmWave communication system with Alamouti scheme.}\label{fig1}\vspace{-5mm}
	\end{center}

\end{figure}
\;\;\;\;We consider a double IRS-aided MU-MISO mmWave communication system in the downlink, where the system uses a hybrid analog/digital precoding structure and the Alamouti scheme, as depicted in Figure 1. The BS is equipped with a uniform linear array (ULA) of $M$ antennas, and $N_{\text{RF}}$ RF chains communicate with a single-user with $K$ antennas or $K$ user equipments (UEs) through 2 IRSs with $R_i$ reflecting elements arranged in a uniform planar array (UPA), where $K \leq N_{\text{RF}} \ll M$. Since the Alamouti scheme is exploited, the number of transmitted symbols should be $K=2$, and $\mathbf{s} \in \mathds{C}^{2\times1}$ is the transmitted symbol vector, which satisfies $\mathds{E}[\mathbf{s}\mathbf{s}^H ]=\frac{P_t}{2} \mathbf{I}_2$, where $P_t$ is the total transmit power. Also, due to the use of the Alamouti encoder at the BS, $T=2$ time slots are used. In this case, the Alamouti transmission matrix would be:
\begin{equation}\label{eq1}
\mathbf{S} = \begin{bmatrix} s_1 & -s_2^* \\ s_2 & s_1^* \end{bmatrix}.
\end{equation}
Throughout transmission, the BS first applies the Alamouti encoder and then passes the encoded information to the digital precoder $\mathbf{F}_{\text{BB}} \in \mathds{C}^{N_{\text{RF}} \times 2}$, followed by an analog precoder $\mathbf{F}_{\text{RF}} \in \mathds{C}^{M \times N_{\text{RF}}}$, where each RF chain is connected to all antennas through some phase shifters (PSTs). Since $\mathbf{F}_{\text{RF}}$ is implemented with the PSTs, each entry of $\mathbf{F}_{\text{RF}}$ satisfies the unit-modulus constraint \cite{ref3}, i.e., $\mathbf{F}_{\text{RF}} \in \mathcal{F} \equiv \left\{ \mathbf{F}_{\text{RF}} \mid \left|[\mathbf{F}_{\text{RF}}]_{m,n}\right|=1, \forall m,n \right\}$. Also, $\mathbf{\Phi}_i = \text{diag}(\tilde{\phi}_i^1, \dots, \tilde{\phi}_i^{R_i}) = \text{diag} \mathbf{(\tilde{\phi}_i)} \in \mathds{C}^{R_i \times R_i}$ is the PS matrix of IRS-$i$, where each $\tilde{\phi}_i^r$ satisfies: $\mathbf{\Phi}_i \in \mathcal{R}_i \equiv \left\{ \tilde{\phi}_i^r \mid \left|\tilde{\phi}_i^r\right|=1, \forall r, i=1,2 \right\}$. Assuming that the direct paths between the BS and the intended UEs are blocked, the received signals at both UEs during two time slots can be written in matrix form as follows:
\begin{equation}\label{eq2}
\mathbf{R} = \left( \mathbf{H}_{I_1} \mathbf{\Phi}_1 \mathbf{H}_{B_1} + \mathbf{H}_{I_2} \mathbf{\Phi}_2 \mathbf{H}_{B_2} \right) \mathbf{F}_{\text{RF}} \mathbf{F}_{\text{BB}} \mathbf{S} + \mathbf{N}
\end{equation}
where the $i^{th}$ row and the $j^{th}$ column of matrix $\mathbf{R}$ for $i,j\in\{1,2\}$ denote the received signal at the $i^{th}$ receiver in the $j^{th}$ time slot. Moreover, $\mathbf{N}$ represents the additive white Gaussian noise matrix, where $n_{ij} \sim \mathcal{CN}(0,\sigma^2)$ for $i,j\in\{1,2\}$. Additionally, $\mathbf{H}_{B_i} \in \mathds{C}^{R_i \times M}$ is the mmWave channel matrix between the IRS-$i$ and the BS, and $\mathbf{H}_{I_i} \in \mathds{C}^{2 \times R_i}$ is the mmWave channel matrix between the $k$-th UE and the IRS-$i$. The mmWave signal has a limited ability to diffract around obstacles due to its small wavelength; therefore, mmWave channels are usually characterized by the so-called extended Saleh-Valenzuela model with $L_B$ and $L_I$ propagation paths for the BS-IRS links and the IRS-UE links, respectively~\cite{ref21, ref22, ref23}. For instance, $\mathbf{H}_{B_1}$ is given by:
\begin{equation}\label{eq3}
\mathbf{H}_{B_1} = \sum_{q=1}^{L_B} \alpha_q \mathbf{a}_r (\psi_q^r, \beta_q^r) \mathbf{a}_t (\psi_q^t, \beta_q^t)^H
\end{equation}
where $\alpha_q$ is the gain of the $q$-th path in $\mathbf{H}_{B_1}$, and we have assumed that $\alpha_q$ for $q\in\{1,L_B\}$ are independently distributed with $\mathcal{C}\mathcal{N}(0, \kappa^2 10^{-0.1 \text{PL}(D)})$, where $\kappa = \sqrt{R_i M / L_B}$ is the normalization factor, and $\text{PL}(D)$ is the path loss that depends on the distance $D$ between the two entities associated with $\mathbf{H}_{B_1}$ \cite{ref23}. Besides, the array response vectors associated with the $q$-th path in $\mathbf{H}_{B_1}$ are respectively denoted by $\mathbf{a}_r (\psi_q^r, \beta_q^r)$ and $\mathbf{a}_t (\psi_q^t, \beta_q^t)$, where $\psi_q^r (\beta_q^r)$ and $\psi_q^t (\beta_q^t)$ represent the azimuth and elevation angles of arrivals and departures (AoAs and AoDs) of the path, respectively. Since it is assumed that UPAs are employed at all these nodes, the transmit array response vectors $\mathbf{a}_{t(r)} (\psi_q^{t(r)}, \beta_q^{t(r)}) $ corresponding to the $q$-th path in $\mathbf{H}_{B_1}$ are given as:
\begin{equation}\label{eq4}
\mathbf{a}_{t(r)} (\psi_q^{t(r)}, \beta_q^{t(r)}) = \frac{1}{\sqrt{M}} 
\begin{bmatrix}
1, & \dots, & e^{j \frac{2\pi d}{\lambda} (m \sin(\psi_q^{t(r)}) \sin(\beta_q^{t(r)}) + n \cos(\beta_q^{t(r)}))}, 
& \dots, 
e^{j \frac{2\pi d}{\lambda} ((W-1) \sin(\psi_q^{t(r)}) \sin(\beta_q^{t(r)}) + (H-1) \cos(\beta_q^{t(r)})} 
\end{bmatrix}^T
\end{equation}
where $\lambda$ is the signal wavelength, $d$ is the distance between the antennas or IRS elements, $0 \leq m \leq W$ and $0 \leq n \leq H$ denote the horizontal and vertical antenna element indices of the node, respectively. Besides, the whole node size is $L = WH$ \cite{ref23}. Other array response vectors resulting in the channel matrices can be similarly defined.

\section{Problem Formulation}
In this paper, we aim to jointly design the HP precoders at the BS and PSs at the IRSs to maximize the SNR as we have utilized the Alamouti Scheme. For simplicity, we define a new channel matrix $\mathbf{G}$ as below,
\begin{equation}\label{eq5}
\mathbf{F} = \mathbf{F}_{\text{RF}} \mathbf{F}_{\text{BB}} \in \mathds{C}^{M \times 2}\;\;\;\;\;
\end{equation}
\begin{equation}\label{eq6}
\mathbf{H}_i = \mathbf{H}_{I_i} \mathbf{\Phi}_i \mathbf{H}_{B_i} \in \mathds{C}^{2 \times M}
\end{equation}
\begin{equation}\label{eq7}
\mathbf{G} = \mathbf{H}_1 \mathbf{F} + \mathbf{H}_2 \mathbf{F} \in \mathds{C}^{2 \times 2}
\end{equation}
Therefore, equation (\ref{eq2}) can equivalently be written by elements of matrices, according to equation (\ref{eq7}),
\begin{equation}\label{eq8}
\begin{bmatrix}
r_{11} & r_{12} \\
r_{21} & r_{22}
\end{bmatrix}
=
\begin{bmatrix}
g_{11} & g_{12} \\
g_{21} & g_{22}
\end{bmatrix}
\begin{bmatrix}
s_1 & -s_2^* \\
s_2 & s_1^*
\end{bmatrix}
+
\begin{bmatrix}
n_{11} & n_{12} \\
n_{21} & n_{22}
\end{bmatrix}
\end{equation}
The procedure to calculate the SNR using the Alamouti scheme with $K=2$ transmit symbols and a single user with 2 receive antennas, can be written in the form of equation (\ref{A2}):
\begin{equation}\label{A2}
\begin{aligned}
& T_1: r_{i1} = g_{i1} s_1 + g_{i2} s_2 + n_{i1} \\
& T_2: r_{i2} = -g_{i1} s_2^* + g_{i2} s_1^* + n_{i2}
\end{aligned}, \quad i=1,2
\end{equation}
where $g_{ij}$ represents the channel matrix value between $j^{th}$ timeslot and $i^{th}$ receiver. Additionally, the Gaussian noise at $i^{th}$ receiver antenna  and $j^{th}$ timeslot is denoted as $n_{ij}$. Now, the above equations can be written in vector form as shown below,
\begin{equation}\label{A3}
\begin{aligned}
& T_1: \mathbf{r}_1 = \mathbf{g}_1 s_1 + \mathbf{g}_2 s_2 + \mathbf{n}_1 \\
& T_2: \mathbf{r}_2 = -\mathbf{g}_1 s_2^* + \mathbf{g}_2 s_1^* + \mathbf{n}_2
\end{aligned}
\end{equation}
where the channel matrix value at all receivers and $j^{th}$ time slot is denoted as $\mathbf{g}_j$. Note that an arbitrary number of receive antennas can be employes; however, for ease of problem folmulation, we have simply considered $n_R=2$ antennas. In this case, with the advantage of using the Alamouti scheme and incorporating the maximum ratio combining (MRC) at the receiver, the Maximum Likelihood (ML) estimate of the received symbol $s_1$, i.e., $\hat{s}_1$, can be written as follow,
\begin{equation}\label{A4}
\hat{s}_1 = s_1 + \frac{\mathbf{g}_1^H \mathbf{n}_1 + \mathbf{n}_2^H \mathbf{g}_2}{|\mathbf{g}_1|^2 + |\mathbf{g}_2|^2} = s_1 + \frac{\hat{n}_1}{|\mathbf{g}_1|^2 + |\mathbf{g}_2|^2}
\end{equation}
Also, by the same token, the ML estimate of $s_2$, i.e., $\hat{s}_2$, can be readily extracted. In this case, the SNR for symbols $s_1$ and $s_2$ can be calculated as follow,
\begin{equation}\label{A5}
\text{SNR}(s_1) = \text{SNR}(s_2) = \frac{P_t}{2\sigma^2} \text{Tr}(\mathbf{G}\mathbf{G}^H) = \frac{P_t}{2\sigma^2} \|\mathbf{G}\|_F^2
\end{equation}
From equation (\ref{A5}), it is inferred that maximizing the symbol SNR of an IRS-aided Alamouti scheme, is equivalent to maximizing the Frobenius norm of the equivalent channel matrix as follow,
\begin{equation}\label{A6}
\max \text{SNR} \equiv \max_{\mathbf{G}} \text{Tr}(\mathbf{G}\mathbf{G}^H)
\end{equation}
 This work can be readily extended to the case of two single-antenna users operating in the low SNR region. In such case, the achievable rate for each user can be approximated as $\log(1+\text{SNR})\approx \text{SNR}$. Therefore, maximizing the total sum-rate is equivalent to maximizing the sum of the SNR values. As shown in Appendix~\ref{Appendix1}, this leads to $\max_{\mathbf{G}} \text{Tr}(\mathbf{G}\mathbf{G}^H)$. As a result, since maximizing the achievable rate of the Alamouti scheme is equivalent to maximizing the overall SNR of each transmitted symbol, we propose the following optimization problem:  

\begin{equation}\label{eq13}
\begin{aligned}
\mathcal{P}1 &: \max_{\mathbf{\Phi}_1, \mathbf{\Phi}_2, \mathbf{F}_{\text{RF}}, \mathbf{F}_{\text{BB}}} \|\mathbf{G}\|_F^2 \\
& \text{s.t.} \\
& \mathbf{C_1}: \mathbf{F}_{\text{RF}} \in \mathcal{F} \equiv \left\{ \mathbf{F}_{\text{RF}} \mid \left|[\mathbf{F}_{\text{RF}}]_{m,n}\right| = 1, \forall m,n \right\} \\
& \mathbf{C_2}: \mathbf{\Phi}_i \in \mathcal{R}_i \equiv \left\{ \tilde{\phi}_i^r \mid \left|\tilde{\phi}_i^r \right| = 1, \forall r, i=1,2 \right\} \\
& \mathbf{C_3}: \|\mathbf{F}_{\text{RF}} \mathbf{F}_{\text{BB}}\|_F^2 = 2
\end{aligned}
\end{equation}
where $\mathbf{C_1}$ and $\mathbf{C_2}$ are, respectively, non-convex unit-modulus constraints associated with the analog precoder and PSs of IRSs. Also, $\mathbf{C_3}$ comes from the power constraint at the BS~\cite{ref5}. Since the optimization variables are coupled in constraint $\mathbf{C_3}$ and the objective function of $\mathcal{P}1$, solving (\ref{eq13}) is inherently difficult to deal with. To tackle the problem, new variables are defined as follows:
\begin{equation}\label{eq14}
\mathbf{H}_B = 
\begin{bmatrix}
\mathbf{H}_{B_1} \\
\mathbf{H}_{B_2} 
\end{bmatrix} \in \mathds{C}^{2R \times M},
\quad
\mathbf{H}_I = 
\begin{bmatrix}
\mathbf{H}_{I_1} & \mathbf{H}_{I_2} 
\end{bmatrix} \in \mathds{C}^{2 \times 2R},
\end{equation}
\begin{equation}\label{eq15}
\mathbf{F} = \mathbf{F}_{\text{RF}} \mathbf{F}_{\text{BB}} \in \mathds{C}^{M \times 2},
\quad
\mathbf{\Phi} = 
\begin{bmatrix}
\mathbf{\Phi}_1 & \mathbf{0} \\
\mathbf{0} & \mathbf{\Phi}_2 
\end{bmatrix} = \text{diag}(\{\tilde{\phi}^1, \dots, \tilde{\phi}^{2R} \}) = \text{diag}(\mathbf{\tilde{\phi}_i}) \in \mathds{C}^{2R \times 2R}
\end{equation}
So, the channel matrix $\mathbf{G}$ can be rewritten as,
\begin{equation}\label{eq16}
\mathbf{G} = (\mathbf{H}_{I_1} \mathbf{\Phi}_1 \mathbf{H}_{B_1} + \mathbf{H}_{I_2} \mathbf{\Phi}_2 \mathbf{H}_{B_2}) \mathbf{F}_{\text{RF}} \mathbf{F}_{\text{BB}} = \mathbf{H}_I \mathbf{\Phi} \mathbf{H}_B \mathbf{F} \in \mathds{C}^{2 \times 2}
\end{equation}
To achieve a simpler solution, the optimization problem is decomposed into two subproblems. In the first subproblem, the new matrices $\mathbf{\Phi}$ and $\mathbf{F}$ are jointly optimized. It should be noted that the constraint $\mathbf{C_1}$ of $\mathcal{P}1$ is not related to this subproblem, hence, it is discarded from constraints associated with this problem. Moreover, constraint $\mathbf{C_2}$ is rewritten based on the newly defined diagonal matrix $\mathbf{\Phi}$ and constraint $\mathbf{C_3}$ is rewritten according to the new matrix $\mathbf{F}$. In the second subproblem, motivated by the work done in~\cite{ref20}, we aim at optimizing analog~($\mathbf{F}_{\text{RF}}$) and digital~($\mathbf{F}_{\text{BB}}$) precoding matrices through minimizing the Euclidean distance between $\mathbf{F}$ and $\mathbf{F}_{\text{RF}} \mathbf{F}_{\text{BB}}$, considering the modulus and power constraints $\mathbf{C_1}$ and $\mathbf{C_3}$ associated with problem $\mathcal{P}1$, respectively. 
It is worth mentioning that the Alamouti scheme is mainly used when CSIT is not available. However, in our work, it is assumed that the CSI associated with BS-to-IRS links as well as IRS-to-user links are available at the transmitter. In such a case, one may argue that transmit beamforming may outperform the Alamouti scheme. However, since the PS matrices associated with IRSs are not known in advance, the use of transmit beamforming may not be feasible. On the other hand, the use of Alamouti code not only achieves the maximum diversity gain, but it also facilitates the decoding procedure without any prior knowledge about PS matrices. In the following, the original problem in $\mathcal{P}1$ is simplified by replacing it with the following two sub-problems.
\section{First Sub-problem}
This section aims at optimizing matrices $\mathbf{\Phi}$ and $\mathbf{F}$. To this end, the problem $\mathcal{P}2$ is formulated as follows,
\begin{equation}\label{eq17}
\begin{aligned}
\mathcal{P}2: & \max_{\mathbf{\Phi}, \mathbf{F}} \|\mathbf{G}\|_F^2 = \|\mathbf{H}_I \mathbf{\Phi} \mathbf{H}_B \mathbf{F}\|_F^2  = \mathcal{Y}_1(\mathbf{\Phi},\mathbf{F})\\
    & \text{s.t.} \\
    & \mathbf{C_1}: \mathbf{\Phi} \in \mathcal{R} \equiv \left\{ \tilde{\phi}^r \mid |\tilde{\phi}^r| = 1, \, r \in [1,2R] \right\} \\
    & \mathbf{C_2}: \|\mathbf{F}\|_F^2 = 2
\end{aligned}
\end{equation}
To tackle the problem, the objective function of $\mathcal{P}2$, i.e. $\mathcal{Y}_1(\mathbf{\Phi},\mathbf{F})$, is rewritten as follows,
\begin{equation}\label{eq18}
\|\mathbf{H}_I \mathbf{\Phi} \mathbf{H}_B \mathbf{F}\|_F^2 = \text{Tr}(\mathbf{H}_I \mathbf{\Phi} \mathbf{H}_B \mathbf{F} (\mathbf{H}_I \mathbf{\Phi} \mathbf{H}_B \mathbf{F})^H) = \text{Tr}((\mathbf{H}_I \mathbf{\Phi})^H \mathbf{H}_I \mathbf{\Phi} \mathbf{H}_B \mathbf{F} (\mathbf{H}_B \mathbf{F})^H)
\end{equation}
Then, defining new matrices $\mathbf{A}=(\mathbf{H}_I \mathbf{\Phi})^H \mathbf{H}_I \mathbf{\Phi}$ and $\mathbf{B}=\mathbf{H}_B \mathbf{F} (\mathbf{H}_B \mathbf{F})^H$, the objective function in $\mathcal{P}2$ becomes $\|\mathbf{G}\|_F^2 =\text{Tr}(\mathbf{A} \mathbf{B})$. Furthermore, matrix $\mathbf{A}$ can be reformulated to form a new optimization matrix $\mathbf{Q}$ which corresponds to the diagonal matrix $\mathbf{\Phi}$ as is detailed in Appendix~\ref{Appendix2}. In this case, we have,
\begin{equation}\label{eq20}
\mathbf{A} = (\mathbf{H}_I \mathbf{\Phi})^H \mathbf{H}_I \mathbf{\Phi} = \mathbf{\Phi}^H \mathbf{H}_I^H \mathbf{H}_I \mathbf{\Phi} = \text{diag}(\mathbf{\tilde{\phi}}^H) \mathbf{H}_I^H \mathbf{H}_I \text{diag}(\mathbf{\tilde{\phi}}) = \mathbf{H}_I^H \mathbf{H}_I \odot \mathbf{\tilde{\phi}} \mathbf{\tilde{\phi}^H} = \mathbf{H}_I^H \mathbf{H}_I \odot \mathbf{Q} = \tilde{\mathbf{H}}_I \odot \mathbf{Q}
\end{equation}
where the auxiliary channel matrix $\tilde{\mathbf{H}}_I$ is defined as $\tilde{\mathbf{H}}_I = \mathbf{H}_I^H \mathbf{H}_I \in \mathds{C}^{2R \times 2R}$. Moreover, the new optimization matrix is set to $\mathbf{Q} = \mathbf{\tilde{\phi}} \mathbf{\tilde{\phi}^H}$ satisfying the constraint $\mathbf{Q} \succeq 0$ and is of rank-1, i.e., $\text{rank}(\mathbf{Q}) = 1$. Similarly, the new matrix $\mathbf{B}$ can be rewritten to form a new optimization matrix $\mathbf{W}$ as follows,
\begin{equation}\label{eq21}
\mathbf{B} = \mathbf{H}_B \mathbf{F}(\mathbf{H}_B \mathbf{F})^H = \mathbf{H}_B \mathbf{F} \mathbf{F}^H \mathbf{H}_B^H = \mathbf{H}_B \mathbf{W} \mathbf{H}_B^H
\end{equation}
where in~(\ref{eq21}), $\mathbf{W}$ is defined as $\mathbf{W} = \mathbf{F} \mathbf{F}^H$, satisfying $\mathbf{W} \succeq 0$ with the rank constraint, i.e., $\text{rank}(\mathbf{W}) = \min(M,2)$. Now, problem $\mathcal{P}2$ can be reformulated based on the new optimization matrices as follows,
\begin{equation}\label{eq22}
\begin{aligned}
\mathcal{P}3: & \max_{\mathbf{Q}, \mathbf{W}} \text{Tr}((\tilde{\mathbf{H}}_I \odot \mathbf{Q}) (\mathbf{H}_B \mathbf{W} \mathbf{H}_B^H)) = \text{Tr}(\mathbf{A} \mathbf{B}) = \mathcal{Y}_2(\mathbf{Q}, \mathbf{W}) \\
\text{s.t.} & \\
& \mathbf{C_1}: \mathbf{W} \succeq 0, \, \text{rank}(\mathbf{W}) = 2 \\
& \mathbf{C_2}: \mathbf{Q} \succeq 0, \, \text{rank}(\mathbf{Q}) = 1 \\
& \mathbf{C_3}: \text{Tr}(\mathbf{W}) = 2 \\
& \mathbf{C_4}: \text{diagC}(\mathbf{Q}) = \mathbf{1}_{2R \times 1}
\end{aligned}
\end{equation}
As is shown in (\ref{eq22}), unknown matrices $\mathbf{Q}$ and $\mathbf{W}$ are multiplied in the objective function; hence, it is a difficult task to optimize this non-linear objective function. However, one can go further and decouple the multiplication of optimization matrices using the notion of inner approximation approach, as is addressed in \cite{ref24} and \cite{ref25}. In fact, the inner approximation approach can be used to rewrite the term $\text{Tr}(\mathbf{A} \mathbf{B})$ as follows,
\begin{equation}\label{eq23}
\text{Tr}(\mathbf{A} \mathbf{B}) = -\frac{1}{2} \|\mathbf{A}\|_F^2 - \frac{1}{2} \|\mathbf{B}\|_F^2 + \frac{1}{2} \|\mathbf{A} + \mathbf{B}\|_F^2
\end{equation}
Noting above, we can rewrite the objective function of $\mathcal{P}3$ as below,
\begin{equation}\label{eq24}
\mathcal{Y}_2(\mathbf{Q}, \mathbf{W}) = -\frac{1}{2} \|\mathbf{H}_B \mathbf{W} \mathbf{H}_B^H\|_F^2 - \frac{1}{2} \|\tilde{\mathbf{H}}_I \odot \mathbf{Q}\|_F^2 + \frac{1}{2} \|\tilde{\mathbf{H}}_I \odot \mathbf{Q} + \mathbf{H}_B \mathbf{W} \mathbf{H}_B^H\|_F^2
\end{equation}
Although the use of inner approximation method makes the objective function $\mathcal{Y}_2(\mathbf{Q}, \mathbf{W})$ to become simpler, it still remains non-convex. To tackle the problem, following the same approach as is done in~\cite{ref26}, the so-called iterative majorization-minimization (MM) approach can be incorporated. To this end, inspired by \cite{ref25} and using the first-order taylor series approximation, $\mathcal{S}_1(\mathbf{Q}$, $\mathbf{W}) = \frac{1}{2} \|\tilde{\mathbf{H}}_I \odot \mathbf{Q} + \mathbf{H}_B \mathbf{W} \mathbf{H}_B^H\|_F^2$ can be lower bounded by the following convex functions,
\begin{equation}\label{eq25}
\begin{aligned}
\mathcal{S}_1(\mathbf{Q}, \mathbf{W}) \geq \mathcal{S}_1(\mathbf{Q}^{(i)}, \mathbf{W}^{(i)}) & + \text{Tr}(\nabla_{\mathbf{Q}}^H \mathcal{S}_1(\mathbf{Q}^{(i)}, \mathbf{W}^{(i)}) (\mathbf{Q} - \mathbf{Q}^{(i)}))+ \text{Tr}(\nabla_{\mathbf{W}}^H \mathcal{S}_1(\mathbf{Q}^{(i)}, \mathbf{W}^{(i)}) (\mathbf{W} - \mathbf{W}^{(i)}))\equiv \tilde{\mathcal{S}}_1 (\mathbf{Q}^{(i)}, \mathbf{W}^{(i)})
\end{aligned}
\end{equation}
where $\{\mathbf{Q}^{(i)}, \mathbf{W}^{(i)}\}$ corresponds to the solutions obtained at the $i$-th iteration of the MM approach. The steps toward finding $\tilde{\mathcal{S}}_1 (\mathbf{Q}^{(i)}, \mathbf{W}^{(i)})$ are given in Appendix~\ref{Appendix3}. Now, the optimization problem can be reformulated as a semi-definite programming (SDP) problem, as follows,
\begin{equation}\label{eq26}
\begin{aligned}
\mathcal{P}4: & \min_{\mathbf{Q}, \mathbf{W}} -\tilde{\mathcal{Y}}_2 (\mathbf{Q}, \mathbf{W}) = \frac{1}{2} \|\tilde{\mathbf{H}}_I \odot \mathbf{Q}\|_F^2 + \frac{1}{2} \|\mathbf{H}_B \mathbf{W} \mathbf{H}_B^H\|_F^2 - \tilde{\mathcal{S}}_1 (\mathbf{Q}^{(i)}, \mathbf{W}^{(i)}) \\
\text{s.t.} & \\
& \mathbf{C_1}: \mathbf{W} \succeq 0, \, \text{rank}(\mathbf{W}) = 2 \\
& \mathbf{C_2}: \mathbf{Q} \succeq 0, \, \text{rank}(\mathbf{Q}) = 1 \\
& \mathbf{C_3}: \text{Tr}(\mathbf{W}) = 2 \\
& \mathbf{C_4}: \text{diagC}(\mathbf{Q}) = \mathbf{1}_{2R \times 1}
\end{aligned}
\end{equation}
However, constraints $\text{rank}(\mathbf{Q}) = 1$ and $\text{rank}(\mathbf{W}) = 2$ make the problem non-convex. To tackle this non-convexity, these constraints can be removed and added to the objective function as a penalty term. Following the same approach as is done in~\cite{ref27}, for any Hermitian matrix $\mathbf{Q}$, we have $\text{Tr}(\mathbf{Q}) = \sum_{n=1}^N \lambda_n(\mathbf{Q}) \geq \max_{n} \lambda_n(\mathbf{Q}) = |\mathbf{Q}|_2$. Hence, the non-convex rank-1 constraint can be rewritten as follows,
\begin{equation}\label{eq27}
\text{Tr}(\mathbf{Q}) - \|\mathbf{Q}\|_2 = 0
\end{equation}
where $\|\mathbf{Q}\|_2$ denotes the spectral norm of matrix $\mathbf{Q}$, which is the largest eigenvalue of matrix $\mathbf{Q}$, i.e., $\lambda_{max}$. Therefore, the non-convex constraint term in $\mathbf{C_2}$ can be added as a penalty term to the objective function of problem $\mathcal{P}4$ as follows,
\begin{equation}\label{eq28}
\begin{aligned}
\mathcal{P}4-1: & \max_{\mathbf{Q}, \mathbf{W}} \, \tilde{\mathcal{Y}}_2(\mathbf{Q}, \mathbf{W}) - \frac{1}{\eta_{\mathbf{Q}}} (\text{Tr}(\mathbf{Q}) - \|\mathbf{Q}\|_2) \\
& \text{s.t.} \\
& \mathbf{C_1}: \mathbf{W} \succeq 0, \, \text{rank}(\mathbf{W}) = 2 \\
& \mathbf{C_2}: \mathbf{Q} \succeq 0 \\
& \mathbf{C_3}: \text{Tr}(\mathbf{W}) = 2 \\
& \mathbf{C_4}: \text{diagC}(\mathbf{Q}) = \mathbf{1}_{2R \times 1}
\end{aligned}
\end{equation}
where $\eta_{\mathbf{Q}}$ is a penalty coefficient. When $\frac{1}{\eta_{\mathbf{Q}}} \rightarrow \infty$, the solution to the problem $\mathcal{P}4-1$ satisfies equation (\ref{eq27}), ensuring that matrix $\mathbf{Q}$ becomes rank one. Since $\|\mathbf{Q}\|_2$ is convex w.r.t. $\mathbf{Q}$, its first-order Taylor approximation can be computed as follows~\cite{ref25},
\begin{equation}\label{eq29}
\|\mathbf{Q}\|_2 \geq \text{Tr}(\mathbf{u}_{\max}(\mathbf{Q}^{(i)}) \mathbf{u}_{\max}^H(\mathbf{Q}^{(i)})(\mathbf{Q} - \mathbf{Q}^{(i)})) + \|\mathbf{Q}^{(i)}\|_2 \triangleq {U}_{\text{ub}}^{(i)}(\mathbf{Q})
\end{equation}
where $\mathbf{Q}^{(i)}$ is obtained by the first-order Taylor approximation of  $\|\mathbf{Q}\|_2$ at the $i^{th}$ iteration, and $\mathbf{u}_{\max}(\mathbf{Q}^{(i)})$ is the eigenvector corresponding to the largest eigenvalue of matrix $\mathbf{Q}^{(i)}$. Thus, the non-convex term in the objective function of problem $\mathcal{P}4-1$ can be approximated using ${U}_{\text{ub}}^{(i)}(\mathbf{Q})$. The approximated problem can then be updated iteratively using convex optimization solvers such as CVX, which is incorporated in the current work. Finally, using the first-order Taylor approximation, the penalty term in equation (\ref{eq27}) can be rewritten as follows,
\begin{equation}\label{eq30}
g_1^{(i)}(\mathbf{Q}) \triangleq \text{Tr}(\mathbf{Q}) - {U}_{\text{ub}}^{(i)}(\mathbf{Q}) = \text{Tr}(\mathbf{Q}) - \|\mathbf{Q}^{(i)}\|_2 - \text{Tr}(\mathbf{u}_{\max}(\mathbf{Q}^{(i)}) \mathbf{u}_{\max}^H(\mathbf{Q}^{(i)})(\mathbf{Q} - \mathbf{Q}^{(i)}))
\end{equation}
Finally, the term $g_1^{(i)}(\mathbf{Q})$ is substituted as the new penalty term in problem $\mathcal{P}4-1$. Now, following the same approach as is done in~\cite{ref29} for matrix $\mathbf{Q}$ with any arbitrary rank constraint, the matrix $\mathbf{W}$ with a rank-2 constraint in $\mathcal{P}4-1$ can also be added as a penalty function to the objective function of $\mathcal{P}4-1$. To this end, since in~\cite{ref29} it is stated that a rank-$K$ matrix can be written as the sum of $K$ rank-1 matrices, a suitable solution for the rank-2 constraint can be applied to the constraint $\text{rank}(\mathbf{W}) = 2$. The nuclear norm of matrix $\mathbf{W} \in \mathds{C}^{M \times 2}$ is defined as $\|\mathbf{W}\|_* = \sum_{j=1}^r \lambda_j$, where $r$ is defined as $r = \min(M, 2)$, and $\lambda_j$ denotes the singular values of matrix $\mathbf{W}$, which are sorted in descending order. Also, the expression $\text{Tr}(\mathbf{W}) = \sum_{n=1}^N \lambda_n(\mathbf{W}) \geq \sum_{j=1}^r \lambda_j \triangleq \|\mathbf{W}\|_*$ holds. As a result, the non-convex rank-2 constraint in $\mathbf{C_1}$ can be similarly reformulated using the DC method as follows,
\begin{equation}\label{eq31}
\text{Tr}(\mathbf{W}) - \|\mathbf{W}\|_* = 0
\end{equation}
Finally, the penalty term in (\ref{eq31}) is added to the objective function of problem $\mathcal{P}4-1$ as follows,
\begin{equation}\label{eq32}
\begin{aligned}
\mathcal{P}4-2: & \max_{\mathbf{Q}, \mathbf{W}} \, \tilde{\mathcal{Y}}_2(\mathbf{Q}, \mathbf{W}) - \frac{1}{\eta_{\mathbf{Q}}} g_1^{(i)}(\mathbf{Q}) - \frac{1}{\eta_{\mathbf{W}}} (\text{Tr}(\mathbf{W}) - \|\mathbf{W}\|_*) \\
& \text{s.t.} \\
& \mathbf{C_1}: \mathbf{W} \succeq 0 \\
& \mathbf{C_2}: \mathbf{Q} \succeq 0 \\
& \mathbf{C_3}: \text{Tr}(\mathbf{W}) = 2 \\
& \mathbf{C_4}: \text{diagC}(\mathbf{Q}) = \mathbf{1}_{2R \times 1}
\end{aligned}
\end{equation}
where $\eta_{\mathbf{W}}$ is a penalty coefficient, and again, as $\frac{1}{\eta_{\mathbf{W}}} \rightarrow \infty$, the solution of $\mathcal{P}4-2$ satisfies (\ref{eq31}), thereby ensuring that the resulting $\mathbf{W}$ becomes of rank-2, hence, it is a solution of $\mathcal{P}4$. Since $\|\mathbf{W}\|_*$ is convex w.r.t. $\mathbf{W}$, the following inequality constraint holds for its first-order Taylor approximation,
\begin{equation}\label{eq33}
\|\mathbf{W}\|_* \geq \|\mathbf{W}^{(i)}\|_* + \text{Tr}(\nabla_{\mathbf{W}}^H \|\mathbf{W}^{(i)}\|_*(\mathbf{W} - \mathbf{W}^{(i)}))
\end{equation}
where the gradient of $\|\mathbf{W}^{(i)}\|_*$ w.r.t. $\mathbf{W}$ has yet to be computed. Since a rank-2 matrix can be written as the sum of two rank-1 matrices \cite{ref29}, the gradient of $\|\mathbf{W}^{(i)}\|_*$ w.r.t. $\mathbf{W}$ corresponds to two eigenvectors associated with the singular values of $\mathbf{W}^{(i)}$ as follows,
\begin{equation}\label{eq34}
\nabla \|\mathbf{W}^{(i)}\|_* = \sum_{j=1}^2 \mathbf{u}_j(\mathbf{W}^{(i)}) \mathbf{u}_j^H(\mathbf{W}^{(i)})
\end{equation}
where $\mathbf{u}_j$ denotes the eigenvector corresponding to the $j^{th}$ singular value of matrix $\mathbf{W}^{(i)}$. Therefore, 
\begin{equation}\label{eq35}
\|\mathbf{W}\|_* \geq \|\mathbf{W}^{(i)}\|_* + \text{Tr}\left(\sum_{j=1}^2 \mathbf{u}_j(\mathbf{W}^{(i)}) \mathbf{u}_j^H(\mathbf{W}^{(i)})(\mathbf{W} - \mathbf{W}^{(i)})\right) \triangleq {U}_{\text{ub}}^{(i)}(\mathbf{W})
\end{equation}
Similarly, $\mathbf{W}^{(i)}$ is obtained by the first-order Taylor approximation at the $i^{th}$ iteration. Thus, the non-convex term in the objective function of problem $\mathcal{P}4-2$ can be approximated using ${U}_{\text{ub}}^{(i)}(\mathbf{W})$. Similar to the problem $\mathcal{P}4-1$, the approximated problem can be iteratively updated using the so-called optimization solver, i.e., CVX. Consequently, the term $g_2^{(i)}(\mathbf{W})$ results as follows,
\begin{equation}\label{eq36}
g_2^{(i)}(\mathbf{W}) \triangleq \text{Tr}(\mathbf{W}) - {U}_{\text{ub}}^{(i)}(\mathbf{W}) = \text{Tr}(\mathbf{W}) - \|\mathbf{W}^{(i)}\|_* - \sum_{j=1}^2 \text{Tr}(\mathbf{u}_j(\mathbf{W}^{(i)}) \mathbf{u}_j^H(\mathbf{W}^{(i)})(\mathbf{W} - \mathbf{W}^{(i)}))
\end{equation}
Finally, the first subproblem is reformulated as an SDP problem with a concave objective function and some affine functions, which can be effectively solved using CVX~\cite{ref30}. Additionally, a new penalty coefficient can be defined as $\eta = \min(\eta_{\mathbf{Q}}, \eta_{\mathbf{W}})$ to consider both penalty coefficients $\eta_{\mathbf{Q}}$ and $\eta_{\mathbf{W}}$. Finally, the final optimization problem of the first subproblem is reformulated as follows,
\begin{equation}\label{eq37}
\begin{aligned}
\mathcal{P}5: & \max_{\mathbf{Q}, \mathbf{W}} \, \tilde{\mathcal{Y}}_2(\mathbf{Q}, \mathbf{W}) - \frac{1}{\eta} (g_1^{(i)}(\mathbf{Q}) + g_2^{(i)}(\mathbf{W})) \\
& \text{s.t.} \\
& \mathbf{C_1}: \mathbf{W} \succeq 0 \\
& \mathbf{C_2}: \mathbf{Q} \succeq 0 \\
& \mathbf{C_3}: \text{Tr}(\mathbf{W}) = 2 \\
& \mathbf{C_4}: \text{diagC}(\mathbf{Q}) = \mathbf{1}_{2R \times 1}
\end{aligned}
\end{equation}

\section{Second Subproblem}
The optimization of matrices $\mathbf{W}$ and $\mathbf{Q}$ in the first sub-problem, has led to the optimization of matrices $\mathbf{F}$, $\mathbf{\Phi}_1$, and $\mathbf{\Phi}_2$. Building on this, and motivated by the work done in~\cite{ref20}, we aim at providing a second subproblem to compute matrices $\mathbf{F}_{\text{RF}}$ and $\mathbf{F}_{\text{BB}}$ using the optimized matrix $\mathbf{F}$. To determine the analog and digital precoding matrices in the second subproblem, we aim to minimize the Euclidean distance between the fully digital precoder ($\mathbf{F}$) and the product of the analog and digital precoders ($\mathbf{F}_{\text{RF}} \mathbf{F}_{\text{BB}}$). To this end, the following optimization problem for determining the analog and digital precoding matrices is defined,
\begin{equation}\label{eq38}
\begin{aligned}
\mathcal{P}6: & \min_{\mathbf{F}_{\text{RF}}, \mathbf{F}_{\text{BB}}} \, \|\mathbf{F} - \mathbf{F}_{\text{RF}} \mathbf{F}_{\text{BB}}\|_F^2 \\
& \text{s.t.} \\
& \mathbf{C_1}: \|\mathbf{F}_{\text{RF}} \mathbf{F}_{\text{BB}}\|_F^2 = 2 \\
& \mathbf{C_2}: \mathbf{F}_{\text{RF}} \in \mathcal{F} \equiv \left\{ \mathbf{F}_{\text{RF}} \mid |[\mathbf{F}_{\text{RF}}]_{m,n}| = 1, \forall m,n \right\}
\end{aligned}
\end{equation}
where $\mathbf{F} \in \mathds{C}^{M \times 2}$ is the fully digital precoder. Additionally, constraint C1 represents the power constraint at the transmitter, and the non-convex constraint C2 in problem $\mathcal{P}6$, which was removed in the first subproblem, makes solving problem $\mathcal{P}6$ challenging, with no known solutions \cite{ref22}. To tackle this issue, we follow the same methodology as done in~\cite{ref22}. The hybrid precoding design problem $\mathcal{P}6$ is divided into two separate subproblems to address the analog and digital precoding matrices. The method proposed in~\cite{ref20} is based on an alternating optimization (AO) approach, where the analog and digital precoding matrices are optimized recursively.

\subsection{Digital Precoding Design}
If the value of the analog precoding matrix ($\mathbf{F}_{\text{RF}}$) is known, the digital precoding matrix ($\mathbf{F}_{\text{BB}}$) can be computed by solving the problem $P7$ as follows,
\begin{equation}\label{eq39}
\begin{aligned}
\mathcal{P}7: & \min_{\mathbf{F}_{\text{BB}}} \, \text{Tr}[\mathbf{F} \mathbf{F}^H - \mathbf{F}_{\text{RF}} \mathbf{F}_{\text{BB}} \mathbf{F}^H - \mathbf{F} \mathbf{F}_{\text{BB}}^H \mathbf{F}_{\text{RF}}^H + \mathbf{F}_{\text{RF}} \mathbf{F}_{\text{BB}} \mathbf{F}_{\text{BB}}^H \mathbf{F}_{\text{RF}}^H] \\
& \text{s.t.} \\
& \mathbf{C_1}: \text{Tr}(\mathbf{F}_{\text{BB}}^H \mathbf{F}_{\text{RF}}^H \mathbf{F}_{\text{RF}} \mathbf{F}_{\text{BB}}) = 2
\end{aligned}
\end{equation}
To solve problem $\mathcal{P}7$, its Lagrangian function is defined as,
\begin{equation}\label{eq40}
\mathcal{L} = \text{Tr}[\mathbf{F} \mathbf{F}^H - \mathbf{F}_{\text{RF}} \mathbf{F}_{\text{BB}} \mathbf{F}^H - \mathbf{F} \mathbf{F}_{\text{BB}}^H \mathbf{F}_{\text{RF}}^H + \mathbf{F}_{\text{RF}} \mathbf{F}_{\text{BB}} \mathbf{F}_{\text{BB}}^H \mathbf{F}_{\text{RF}}^H] + \lambda [\text{Tr}(\mathbf{F}_{\text{BB}}^H \mathbf{F}_{\text{RF}}^H \mathbf{F}_{\text{RF}} \mathbf{F}_{\text{BB}}) - 2]
\end{equation}
where $\lambda \in \mathds{R}$ is the Lagrangian multiplier that needs to be determined. By taking the derivative of equation (\ref{eq40}) w.r.t. $\mathbf{F}_{\text{BB}}$ and setting it to zero, the optimal value of $\mathbf{F}_{\text{BB}}$ is obtained as below,

\begin{equation}\label{eq41}
\mathbf{F}_{\text{BB}}^* = \frac{1}{(1 + \lambda)} (\mathbf{F}_{\text{RF}}^H \mathbf{F}_{\text{RF}})^{-1} \mathbf{F}_{\text{RF}}^H \mathbf{F} = \beta \tilde{\mathbf{F}}_{\text{BB}}
\end{equation}
where the coefficient $\beta = \frac{1}{(1 + \lambda)}$ is a normalization factor, and the unnormalized digital precoding matrix is defined as $\tilde{\mathbf{F}}_{\text{BB}} = (\mathbf{F}_{\text{RF}}^H \mathbf{F}_{\text{RF}})^{-1} \mathbf{F}_{\text{RF}}^H \mathbf{F} = \mathbf{F}_{\text{RF}}^\dagger \mathbf{F}$ \cite{ref20}.To determine the value of $\beta$, we set the derivative of the Lagrangian function w.r.t. $\lambda$ to 0, and the calculated $\mathbf{F}_{\text{BB}}^*$ from (\ref{eq41}) is then substituted into the Lagrangian. Finally, the value of $\beta$ is obtained as follows,
\begin{equation}\label{eq42}
\beta = \sqrt{\frac{2}{\text{Tr}(\tilde{\mathbf{F}}_{\text{BB}}^H \mathbf{F}_{\text{RF}}^H \mathbf{F}_{\text{RF}} \tilde{\mathbf{F}}_{\text{BB}})}}
\end{equation}

\subsection{Analog Precoding Design}
In~\cite{ref31}, Lemma 1, it is proved that if the Euclidean distance between the fully digital precoding matrix and the unnormalized hybrid precoding matrix is minimized, the normalization step at the end ensures that the distance between the product of the hybrid precoding matrices and the fully digital precoding matrix remains sufficiently small \cite{ref20}. Therefore, the power constraint $\mathbf{C_1}$ in problem $\mathcal{P}6$ can be temporarily set aside, allowing the focus to be solely on minimizing the Euclidean distance between the fully digital precoding matrix and the unnormalized hybrid precoding matrix. Then, the normalization of the digital precoding matrix can be applied after optimizing the analog precoding matrix and the unnormalized digital precoding matrix. Given an unnormalized digital precoding matrix, the subproblem for designing the analog precoding matrix can be rewritten as follows:
\begin{equation}\label{eq43}
\begin{aligned}
\mathcal{P}8: & \min_{\mathbf{F}_{\text{RF}}} \, \text{Tr}[\mathbf{F}\mathbf{F}^H - \mathbf{F}_{\text{RF}} \tilde{\mathbf{F}}_{\text{BB}} \mathbf{F}^H - \mathbf{F} \tilde{\mathbf{F}}_{\text{BB}}^H \mathbf{F}_{\text{RF}}^H + \mathbf{F}_{\text{RF}} \tilde{\mathbf{F}}_{\text{BB}} \tilde{\mathbf{F}}_{\text{BB}}^H \mathbf{F}_{\text{RF}}^H] \\
& \text{s.t.} \\
& \mathbf{C_1}: \, |[\mathbf{F}_{\text{RF}}]_{m,n}| = 1,\;\; \forall m,n
\end{aligned}
\end{equation}
By removing the term $\mathbf{F}\mathbf{F}^H$, which does not depend on $\mathbf{F}_{\text{RF}}$, the subproblem for analog precoding in $\mathcal{P}8$ can be rewritten as follows,
\begin{equation}\label{eq44}
\begin{aligned}
\mathcal{P}8-1: & \min_{\mathbf{F}_{\text{RF}}} \, \text{Tr}[\mathbf{F}_{\text{RF}} \tilde{\mathbf{F}}_{\text{BB}} \tilde{\mathbf{F}}_{\text{BB}}^H \mathbf{F}_{\text{RF}}^H - \mathbf{F}_{\text{RF}} \tilde{\mathbf{F}}_{\text{BB}} \mathbf{F}^H - \mathbf{F} \tilde{\mathbf{F}}_{\text{BB}}^H \mathbf{F}_{\text{RF}}^H] \\
& \text{s.t.} \\
& \mathbf{C_1}: \, |[\mathbf{F}_{\text{RF}}]_{m,n}| = 1,\;\; \forall m,n
\end{aligned}
\end{equation}
To solve the optimization problem $\mathcal{P}8-1$, an equivalent optimization problem is introduced as follows~\cite{ref20},
\begin{equation}\label{eq45}
\begin{aligned}
\mathcal{P}9: & \min_{\tilde{\mathbf{F}}_{\text{RF}}} \, \text{Tr}[\tilde{\mathbf{F}}_{\text{RF}}^H \mathbf{M} \tilde{\mathbf{F}}_{\text{RF}}] \\
& \text{s.t.} \\
& \mathbf{M} = \begin{bmatrix} \tilde{\mathbf{F}}_{\text{BB}} \tilde{\mathbf{F}}_{\text{BB}}^H  & -\frac{1}{\sqrt{M}} \tilde{\mathbf{F}}_{\text{BB}} \mathbf{F}^H \\ -\frac{1}{\sqrt{M}} \mathbf{F} \tilde{\mathbf{F}}_{\text{BB}}^H & \mathbf{0} \end{bmatrix}, \\
& \tilde{\mathbf{F}}_{\text{RF}} = \begin{bmatrix} \bar{\mathbf{F}}_{\text{RF}}^H \\ \mathbf{I}_M \end{bmatrix}, \\
& |\bar{\mathbf{F}}_{\text{RF}}| = \frac{1}{\sqrt{M}} \mathbf{1}_{M \times N_{\text{RF}}}
\end{aligned}
\end{equation}
Expanding $\mathcal{P}9$ reveals that it is equivalent to $\mathcal{P}8-1$ assuming $\mathbf{F}_{\text{RF}} = \sqrt{M} \bar{\mathbf{F}}_{\text{RF}}$. If the optimal solution to problem $\mathcal{P}9$ is given by $\tilde{\mathbf{F}}_{\text{RF}}^* = \begin{bmatrix} \left(\bar{\mathbf{F}}_{\text{RF}}^*\right)^H \\ \mathbf{I}_M \end{bmatrix}$, then one can readily verify that the optimal solution of $\mathcal{P}8-1$ becomes $\mathbf{F}_{\text{RF}}^* = \sqrt{M} \bar{\mathbf{F}}_{\text{RF}}^*$ \cite{ref20}. Therefore, solving $\mathcal{P}9$ gives the analog precoding matrix. To solve problem $\mathcal{P}9$, the Hermitian matrix $\mathbf{X}$ is defined as, 
\begin{equation}\label{eq46}
\mathbf{X} = \tilde{\mathbf{F}}_{\text{RF}} \tilde{\mathbf{F}}_{\text{RF}}^H = \begin{bmatrix} \bar{\mathbf{F}}_{\text{RF}}^H \bar{\mathbf{F}}_{\text{RF}} & \bar{\mathbf{F}}_{\text{RF}}^H \\ \bar{\mathbf{F}}_{\text{RF}} & \mathbf{I}_M \end{bmatrix}
\end{equation}
Since the $i^{th}$ diagonal element of the matrix $\bar{\mathbf{F}}_{\text{RF}}^H \bar{\mathbf{F}}_{\text{RF}}$ is equal to the sum of the squares of the absolute values of all elements in the $i^{th}$ column of the matrix $\bar{\mathbf{F}}_{\text{RF}}$, it is evident that all diagonal elements of $\bar{\mathbf{F}}_{\text{RF}}^H \bar{\mathbf{F}}_{\text{RF}}$ are equal to one. Consequently, $\text{diagC}(\mathbf{X}) = \mathbf{1}_{m \times 1}$, where $m$ is defined as $m \triangleq (N_{\text{RF}} + M)$. Additionally, the indices $q$ and $r$ are defined as $q \triangleq 1:N_{\text{RF}}$ and $r \triangleq (N_{\text{RF}} + 1):(N_{\text{RF}} + M)$, respectively \cite{ref20}. Therefore, problem $\mathcal{P}9$ can be reformulated as follows,
\begin{equation}\label{eq47}
\begin{aligned}
\mathcal{P}10: & \min_{\mathbf{X}} \, \text{Tr}(\mathbf{M} \mathbf{X}) \\
& \text{s.t.} \\
& \mathbf{C_1}: \, \mathbf{X} \succeq 0 , \, \text{rank}(\mathbf{X}) = M \\
& \mathbf{C_2}: \, \text{diagC}(\mathbf{X}) = \mathbf{1}_{m \times 1} \\
& \mathbf{C_3}: \, |\mathbf{X}_{r,q}| = \frac{1}{\sqrt{M}} \mathbf{1}_{M \times N_{\text{RF}}} = |\mathbf{X}_{r,q}^H| \\
& \mathbf{C_4}: \, |\mathbf{X}_{r,r}| = \mathbf{I}_M
\end{aligned}
\end{equation}
Then, applying convex relaxation strategy as is done in \cite{ref20}, problem $\mathcal{P}10$ can be approximated by the following convex problem,
\begin{equation}\label{eq48}	
\begin{aligned}
\mathcal{P}11: & \min_{\mathbf{X}} \, \text{Tr}(\mathbf{M} \mathbf{X}) \\
& \text{s.t.} \\
& \mathbf{C_1}: \, \mathbf{X} \succeq 0 \\
& \mathbf{C_2}: \, \text{diagC}(\mathbf{X}) = \mathbf{1}_{m \times 1} \\
& \mathbf{C_3}: \, |\mathbf{X}_{r,q}| = \frac{1}{\sqrt{M}} \mathbf{1}_{M \times N_{\text{RF}}} = |\mathbf{X}_{r,q}^H| \\
& \mathbf{C_4}: \, |\mathbf{X}_{r,r}| = \mathbf{I}_M
\end{aligned}
\end{equation}
According to \cite{ref32}, it becomes clear that problem $\mathcal{P}11$, after convex relaxation closely resembles the \textit{PhaseCut} problem, differing only by the additional constraint $\mathbf{C_4}$. In reference \cite{ref20}, the Block Coordinate Descent (BCD) approach is chosen to solve the optimization problem $\mathcal{P}11$ while considering the additional constraint $\mathbf{C_4}$, leading to the approach known as Modified Block Coordinate Descent.
\section{Complexity Analysis}
The computational complexity is divided into two parts. First, the computational complexity of the first sub-problem, i.e., $\mathcal{P}5$, is analyzed. It should be noted that in this case, the two matrices $\mathbf{Q} \in \mathds{C}^{2R \times 2R}$ and $\mathbf{W} \in \mathds{C}^{M \times M}$ are jointly optimized by solving the underlying SDP problem using the so-called Interior Point Method, which is embedded in related packages such as CVX, which is used in the current study. For an SDP problem with $n$ variables and $m$ constraints, the computational complexity per iteration is $O(mn^3 + m^2 n^2 + m^3)$ \cite{ref29}. Thus, it is required to identify the total number of optimization variables and constraints to determine the computational complexity. To this end, note that the unknown matrix $\mathbf{Q}$ has $4R^2$ real variables, and $\mathbf{W}$ has $M^2$ real variables. Thus, the total number of optimization variables is $n = 4R^2 + M^2$. Since $R$, the number of IRS elements, is significantly larger than $M$, the antenna elements, it follows that $n$ is of order $O(R^2)$. 
Also, constraints $\mathbf{C_1}$ and $\mathbf{C_2}$ add two constraints, Constraint $\mathbf{C_3}$ is just one constraint, and constraint $\mathbf{C_4}$, which corresponds to the elements on the main diagonal of the optimization variable $\mathbf{Q}$, adds $2R$ constraints. Thus, the total number of constraints is of order $m = O(R)$. As a result, $m$ is of order $O(\sqrt{n})$. Hence, the computational complexity of the first sub-problem scales as $O(n^{3.5})$\cite{ref29}. Therefore, the computational complexity per iteration for the first subproblem is  $O(R^{7})$.
According to~\cite{ref23}, the computational complexity of the second subproblem per iteration is given by $O(N_{\text{it}}^i (M + N_{\text{RF}} - 1)^2 + M^2 N_{\text{RF}} + 2MN_{\text{RF}} K + (N_{\text{RF}})^2 K).$ Finally, the overall computational complexity per iteration for the optimization problem can be expressed as $O(R^{7} + N_{\text{it}}^i (M + N_{\text{RF}} - 1)^2 + M^2 N_{\text{RF}} + 2MN_{\text{RF}} K + (N_{\text{RF}})^2 K)$.

\section{Simulation Results}
\begin{figure}
	\begin{center}
		\includegraphics[scale=0.1]{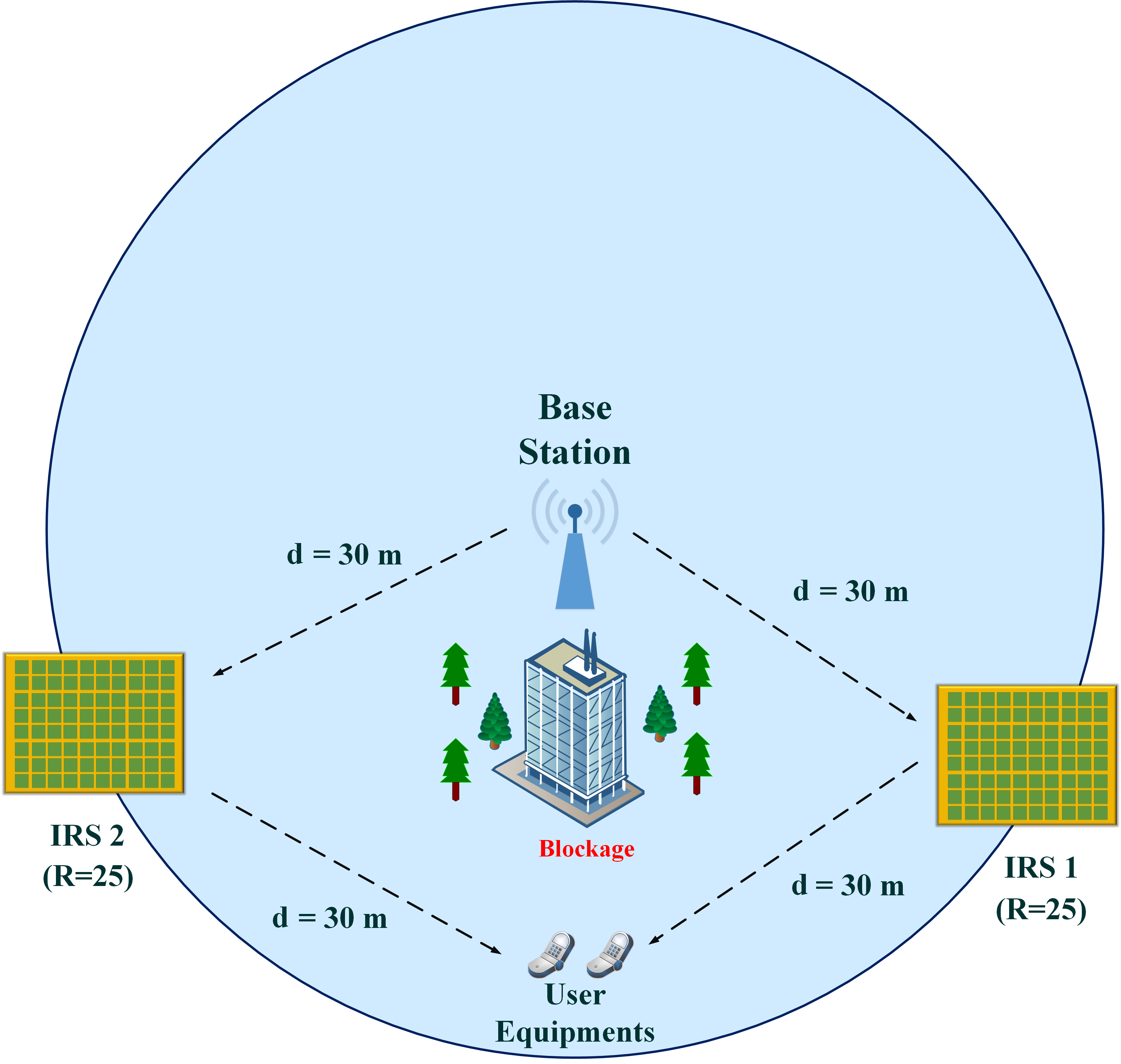}\vspace{-1mm}
		\caption{\;\;Main scenario of double IRS-aided MU-MISO mmWave communication utilizing the Alamouti Scheme.}\label{fig2}\vspace{-5mm}
	\end{center}

\end{figure}

\begin{figure}
	\begin{center}
		\includegraphics[scale=0.8]{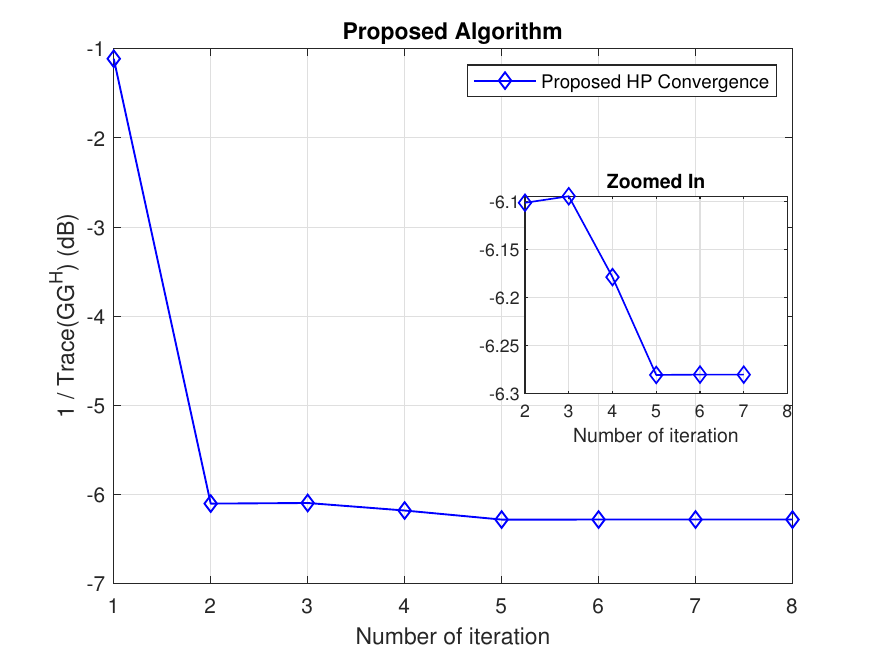}
		\caption{\;\;Convergence of the proposed algorithm.}\label{fig3}\vspace{-5mm}
	\end{center}

\end{figure}

\begin{table}[t!]
\centering
\caption{Simulation Parameters}
\begin{tabular}{|l|l|}
\hline
\textbf{Simulation Parameter} & \textbf{Values} \\ \hline
Number of Transmit antennas & $M=10$ \\ \hline
Number of Users & $K=2$ \\ \hline
Number of RF chains & $N_{\text{RF}}=2$ \\ \hline
Each IRS elements & $R=25$ \\ \hline
Transmit power (dBm) & $P_t=30$ dBm \\ \hline
Transmit antenna gain \cite{ref23} & $g_t=49$ dBi \\ \hline
Paths (BS to each IRS) & $L_B=5$ \\ \hline
Paths (each IRS to users) & $L_I=5$ \\ \hline
Distance between BS and each IRS & $d_{H_B}=30$ m \\ \hline
Distance between each IRS and users & $d_{H_I}=30$ m \\ \hline
\end{tabular}
\label{tab1}
\end{table}
This section investigates the performance of the proposed algorithm for designing hybrid precoders and PSs in both IRSs utilizing the Alamouti scheme through numerical simulations. The main simulation parameters are summarized in Table \ref{tab1} unless stated otherwise. Additionally, the locations of the transmitter, IRSs, and users is illustrated in Figure \ref{fig2}. It is assumed that the direct paths between the transmitter and the users are blocked by obstacles. Therefore, the transmitted information vector at the base station is only reflected to the users through the two IRSs. According to Figure \ref{fig3}, the proposed algorithm converges after four iterations.

\begin{figure}
	\begin{center}
		\includegraphics[scale=0.8]{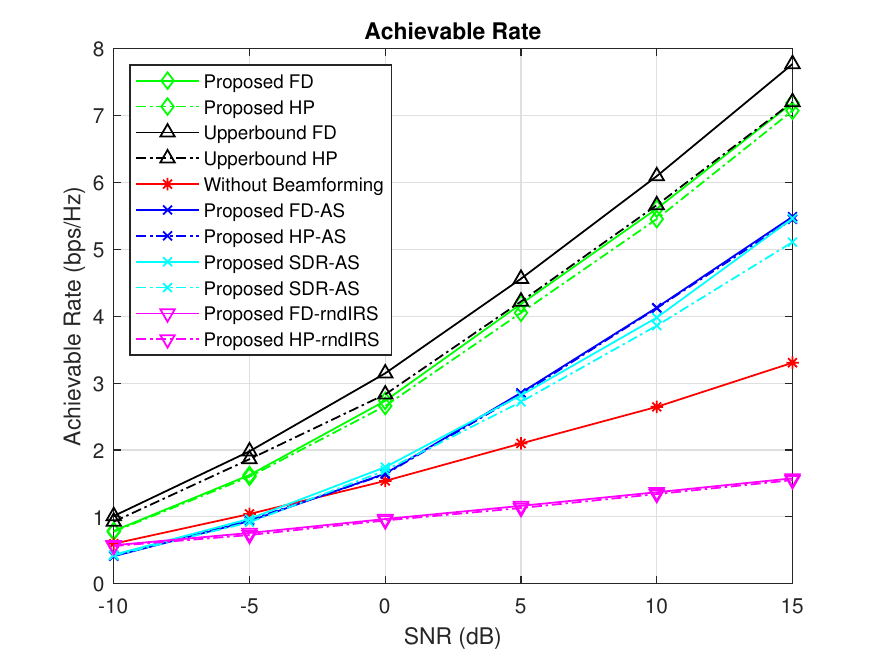}\vspace{-1mm}
		\caption{\;\;Comparison of the Achievable Rate of the Proposed algorithm with benchmark Schemes.}\label{fig4}\vspace{-5mm}
	\end{center}

\end{figure}

\noindent In Figure \ref{fig4}, the performance of the proposed algorithm is compared with several benchmark schemes. The descriptions of the schemes used for comparison are as follows:\\
1) \textbf{FD}: Fully digital precoding is utilized at the BS. 2) \textbf{HP}: Hybrid analog/digital precoding is utilized at the BS. 3) \textbf{UpperBound}: The penalty approach is not used for the rank constraints, and convex relaxation is applied to all rank constraints. 4) \textbf{Without Beamforming}: No precoding is utilized at the transmitter. 5) \textbf{AS(Antenna Selection)}: Only the two antennas with the highest channel weights are used for transmitting information. 6) \textbf{rndIRS}: The PSs are set randomly. 7) \textbf{SDR}: The scheme in which the convex relaxation approach is applied is denoted as SDR. \\
Note that in all the above schemes, the rest of the proposed algorithm is used to optimize the remaining variables. Based on the above descriptions and Figure \ref{fig4}, it can be concluded that the proposed algorithm performs well compared to the upper bound. When hybrid precoding is used at the BS, the performance difference between \textit{Proposed HP} and \textit{UpperBound HP} in the SNR range of 5 to 15 dB is less than 4\%. Moreover, at an SNR of 5 dB, the rate difference improves up to 0.17 bps/Hz. Additionally, when fully digital precoding is used at the BS, the performance difference between \textit{Proposed FD} and \textit{UpperBound FD} in the SNR range of 5 to 15 dB is less than 8\%, with a rate difference of 0.3 bps/Hz at an SNR of 5 dB. Comparing \textit{Proposed HP} with \textit{UpperBound FD}, we observe that their performance difference in the SNR range of 5 to 15 dB is less than 10\%, with a rate difference of 0.5 bps/Hz at an SNR of 5 dB. Furthermore, the proposed algorithm significantly outperforms the other benchmark schemes. \\
\begin{figure}
	\begin{center}
		\includegraphics[scale=0.8]{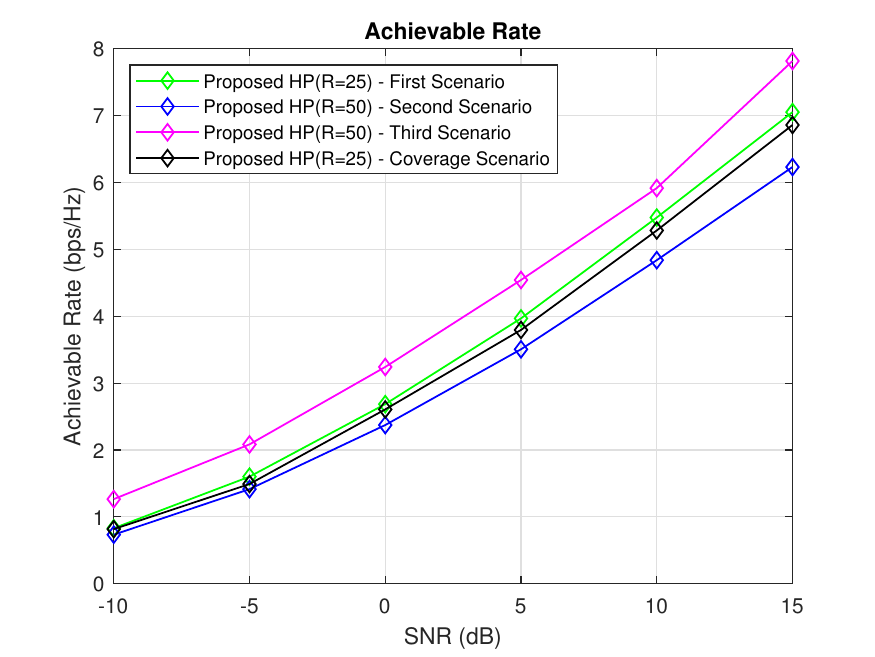}\vspace{-1mm}
		\caption{\;\;Performance Analysis of the Proposed Scheme with Increasing Reflecting Elements of IRSs and Changing User locations.}\label{fig5}\vspace{-5mm}
	\end{center}

\end{figure}

\noindent To analyze Figure \ref{fig5}, it is first necessary to review the scenarios depicted in this figure. The descriptions of the various scenarios are as follows:\\
1) \textbf{First Scenario}: The distance between the transmitter and the IRSs, as well as the distance between the IRSs and the users, is set to 30 meters. Additionally, the number of reflecting elements per IRS is 25. The users are located close to each other. 2) \textbf{Second Scenario}: The distance between the transmitter and the first IRS, as well as the distance between the first IRS and the users, is set to 30 meters, while these distances are set to 600 meters for the second IRS. Moreover, the number of reflecting elements per IRS is 50. The users are located close to each other. 3) \textbf{Third Scenario}: The distance between the transmitter and the IRSs, as well as the distance between the IRSs and the users, is set to 30 meters. Additionally, the number of reflecting elements per IRS is 50. The users are located close to each other. 4) \textbf{Coverage Scenario}: The distance between the transmitter and the IRSs is 30 meters, while the first user is 15 meters away from the first IRS and 50 meters away from the second IRS. Similarly, the second user is 15 meters away from the second IRS and 50 meters away from the first IRS. The number of reflecting elements per IRS is 25.\\
Comparing the \textit{First Scenario} and the \textit{Second Scenario}, we conclude that when the second IRS is relatively far from the transmitter and the users, the transmitted information experiences severe attenuation. Consequently, the users mainly receive information from the reflected path of the first IRS. Even if the number of reflecting elements in the IRSs is doubled, the information from the reflected path of the second IRS will still experience significant attenuation. Therefore, the number of effective reflecting elements in the IRSs can be considered nearly equal in both scenarios. As a result, the performance of the \textit{First Scenario} compared to the \textit{Second Scenario} is similar to comparing the performance of using one IRS versus two IRSs. Hence, the \textit{First Scenario} offers better performance than the \textit{Second Scenario}, as shown in Figure \ref{fig5}. From the comparison between the \textit{First Scenario} and the \textit{Third Scenario}, we conclude that if the locations of the transmitter, IRSs, and users are fixed, and only the number of reflecting elements in the IRSs is doubled, the performance of the communication network improves. Specifically, the performance of the \textit{Third Scenario} compared to the \textit{First Scenario} can improve by up to 0.6 bps/Hz at an SNR of 5 dB. Comparing the \textit{First Scenario} with the \textit{Coverage Scenario}, we conclude that if the locations of the transmitter and IRSs are fixed, and each user is located near only one IRS, the performance of the \textit{Coverage Scenario} slightly reduces compared to the \textit{First Scenario}. However, the \textit{Coverage Scenario} shows a significant performance improvement compared to the \textit{Second Scenario}, which is similar to using one IRS. Thus, it can be concluded that increasing the number of IRSs leads to a significant improvement in the coverage of the IRS-aided communication network.\\
\begin{figure}
	\begin{center}
		\includegraphics[scale=0.8]{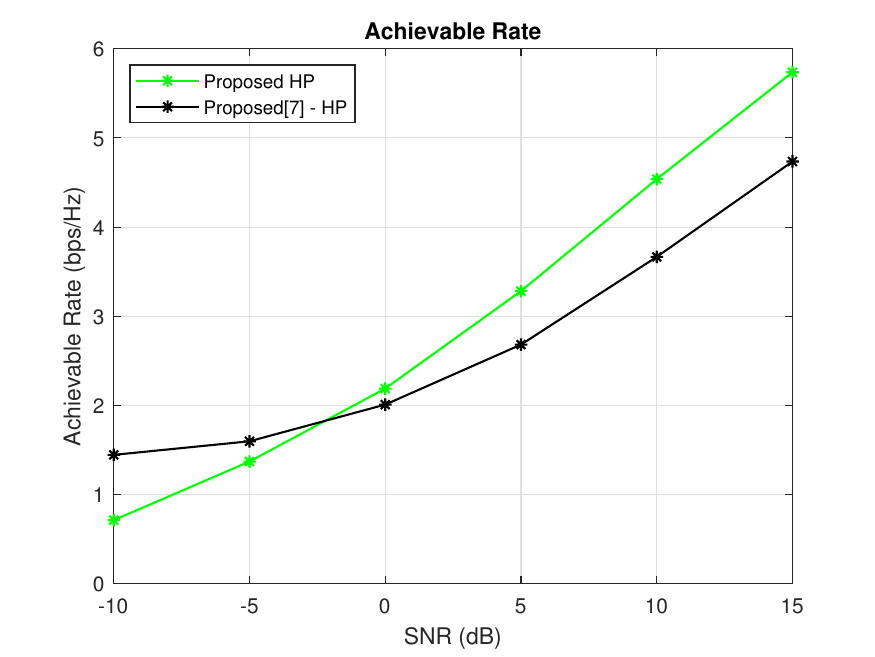}\vspace{-1mm}
		\caption{\;\;Comparing Proposed algorithm with method presented in reference \cite{ref7}.}\label{fig6}\vspace{-5mm}
	\end{center}

\end{figure}

\begin{figure}
	\begin{center}
		\includegraphics[scale=0.8]{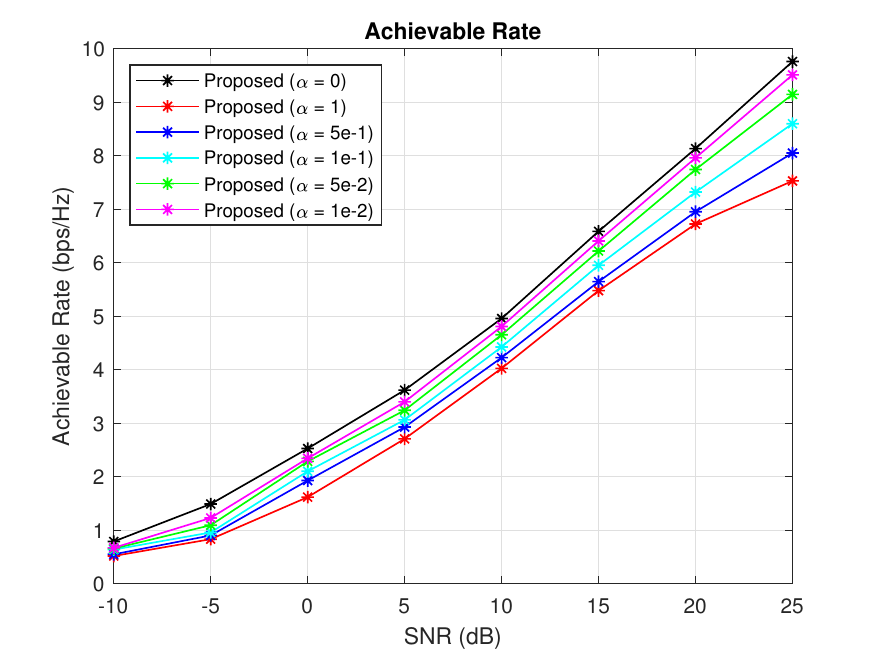}\vspace{-1mm}
		\caption{\;\;Comparison of the Impact of Millimeter-Wave Channel Variations.}\label{fig7}\vspace{-5mm}
	\end{center}

\end{figure}

\noindent In Figure \ref{fig6}, the performance of the proposed scheme (\textit{Proposed HP}) is compared with the approach presented in~\cite{ref7} (\textit{Proposed[7] - HP}). For both approaches, a single IRS is considered, and other simulation parameters are set as described in Table~\ref{tab1}. An alternating optimization approach is used to configure the analog precoder matrix, the digital precoder matrix, and the PSs of the IRS in~\cite{ref7}, whereas our proposed scheme employs a joint optimization approach. Additionally, the Alamouti coding scheme is not utilized in~\cite{ref7}, requiring the calculation of the Signal-to-Interference-and-Noise Ratio (SINR) for that approach. In contrast, only the SNR is calculated in our proposed algorithm. Due to the use of Alamouti coding, note that a constant factor of 0.5 is considered in the rate calculation for our scheme. Based on Figure \ref{fig6}, it can be concluded that our proposed method performs better at higher SNRs compared to the approach in~\cite{ref7}, due to the significant impact of interference. It is worth mentioning that the performance improvement of our proposed method, in comparison to the approach in reference \cite{ref7}, exceeds 21\% in the SNR range of 5 to 15 dB. \\ 

\noindent In Figure \ref{fig7}, the performance of the proposed communication network is examined for various uncertainty values. The uncertainty is modeled by adding a normally distributed random matrix, defined as $\Delta\mathbf{H} \sim \mathcal{CN}(0,1)$, to the channel matrices $\mathbf{H}_{B_1}$, $\mathbf{H}_{I_1}$, $\mathbf{H}_{B_2}$ and $\mathbf{H}_{I_2}$ with a scaling factor $\alpha$. For instance, the new matrix $\mathbf{H}_{B_1}^{(\text{new})}$ is formed as $\mathbf{H}_{B_1}^{(\text{new})} = \mathbf{H}_{B_1} + \alpha \Delta\mathbf{H}$. From Figure \ref{fig7}, it can be inferred that increasing $\alpha$ enhances the interference term, which in turn reduces the upward trend of the curves in the high SNR region. For example, when $\alpha = 1$, the achievable rate shows a noticeable reduction in the upward trend for values above 20 dB, indicating the impact of interference and resulting in a break point.

\section{Conclusion}
This paper focuses on the joint optimization of hybrid precoding and IRS beamforming matrices in mmWave MU-MISO Multi IRS-aided communication networks, utilizing the Alamouti scheme to maximize the signal-to-noise ratio. The approach addresses the non-convex optimization problem by reformulating it into two sub-problems, which are solved using inner approximation, majorization-minimization, and a modified block coordinate descent algorithm. The Alamouti scheme simplifies the optimization process, requiring only two coefficients for decoding at each user. Simulation results demonstrate that the proposed algorithm outperforms benchmark schemes in terms of achievable rate, highlighting the effectiveness of the joint hybrid precoding design. The integration of the Alamouti coding scheme in the proposed algorithm yields performance close to the fully digital case. This paper's results demonstrates that employing two IRSs significantly improves coverage in millimeter-wave communication networks, offering a practical and cost-effective solution to address challenges such as propagation loss and blockage in mmWave communication.

\bibliographystyle{IEEEtran}
\bibliography{references}

\appendix
\subsection{Calculate SNR of 2 Single User antenna}\label{Appendix1}
In this scenario, two single-antenna users are assumed, and the goal is to maximize the sum of the SNRs of both users, with the MRC receiver applied to calculate the received signals for each user. According to equation (\ref{eq8}), the approach proceeds as follows for the received signal of the first user: The complex conjugate of $r_{12}$ is multiplied by $g_{12}$, and $r_{11}$ is multiplied by $g_{11}^*$. Then, both values are summed together. Finally, the received signal for the first user is obtained as follows:
\begin{equation}\label{eq9}
\hat{r}_1 = \left( |g_{11}|^2 + |g_{12}|^2 \right)s_1  + \hat{n}_1
\end{equation}
Similarly, the received signal for the second user is obtained as follows,
\begin{equation}\label{eq10}
\hat{r}_2 = \left( |g_{21}|^2 + |g_{22}|^2 \right)s_2 + \hat{n}_2
\end{equation}
Now, the SNR of each user can be computed as:
\begin{equation}\label{eq11}
\text{SNR}(s_i) = \frac{P_t}{2\sigma^2} \left( |g_{i1}|^2 + |g_{i2}|^2 \right)
\end{equation}
Then, maximizing the sum of SNRs is equivalent to maximizing $\text{Tr}(\mathbf{G}\mathbf{G}^H)$.
\begin{equation}\label{eq12}
\begin{aligned}
& \max \text{SNR}_{\text{total}} = \max \left( \frac{P_t}{2\sigma^2} \sum_{i=1}^2 \left( |g_{i1}|^2 + |g_{i2}|^2 \right) \right) \\
& \quad \quad \quad \quad \quad \;\; = \max_{\mathbf{G}} \text{Tr}(\mathbf{G}\mathbf{G}^H)
\end{aligned}
\end{equation}
In this case, the final channel matrix $\mathbf{G}$ is only partially available to the users' receivers. Instead of maximizing the individual SNRs, the sum of the SNRs is maximized, which approximates the sum rate maximization at low SNR levels. A key advantage of the proposed method is that user $i$ only needs the $i^{th}$ row of matrix $\mathbf{G}$. For instance, user 1 requires only $g_{11}^*$ and $g_{12}$ for decoding. Therefore, each user needs only two coefficients, regardless of the number of antennas at the transmitter and IRS elements. These coefficients can be computed by a central node and sent to each user via a control channel. Hence, the decoding process at the receiver is independent of the number of transmit antennas and IRS elements.

\subsection{Proof of $\mathbf{X}^H \mathbf{A} \mathbf{X} = \mathbf{A} \odot (\mathbf{x} \mathbf{x}^H)$}\label{Appendix2}
Assume that the diagonal matrix $\mathbf{X}$ is defined as $\mathbf{X} = \text{diag}(\mathbf{x})$, where $\mathbf{x} \in \mathds{C}^{n \times 1}$. Moreover, let $\mathbf{A} \in \mathds{C}^{n \times n}$ be a square matrix. We now need to verify the validity of equation (\ref{B1}) as follows:
\begin{equation}\label{B1}
\mathbf{X}^H \mathbf{A} \mathbf{X} = \mathbf{A} \odot (\mathbf{x} \mathbf{x}^H)
\end{equation}
To do this, we first rewrite the left-hand side of the equality in matrix form, as shown in~(\ref{B2}),
\begin{equation}\label{B2}
\begin{aligned}
& \mathbf{X}^H \mathbf{A} \mathbf{X} = 
\begin{bmatrix}
x_1^* & \cdots & 0 \\
\vdots & \ddots & \vdots \\
0 & \cdots & x_n^*
\end{bmatrix}
\begin{bmatrix}
a_{11} & \cdots & a_{1n} \\
\vdots & \ddots & \vdots \\
a_{n1} & \cdots & a_{nn}
\end{bmatrix}
\begin{bmatrix}
x_1 & \cdots & 0 \\
\vdots & \ddots & \vdots \\
0 & \cdots & x_n
\end{bmatrix} \\
& \quad \quad \quad \; =
\begin{bmatrix}
a_{11} x_1 x_1^* & \cdots & a_{1n} x_n x_1^* \\
\vdots & \ddots & \vdots \\
a_{n1} x_1 x_n^* & \cdots & a_{nn} x_n x_n^*
\end{bmatrix}
\end{aligned}
\end{equation}
Next, to verify equation (\ref{B1}), the right-hand side of the equality is rewritten in matrix form, as shown in~(\ref{B3}),
\begin{equation}\label{B3}
\begin{aligned}
& \mathbf{A} \odot (\mathbf{x} \mathbf{x}^H) =
\begin{bmatrix}
a_{11} & \cdots & a_{1n} \\
\vdots & \ddots & \vdots \\
a_{n1} & \cdots & a_{nn}
\end{bmatrix}
\odot
\begin{bmatrix}
x_1 \\
\vdots \\
x_n
\end{bmatrix}
\begin{bmatrix}
x_1^* & \cdots & x_n^*
\end{bmatrix}\\
& \quad \quad \quad \quad \;\;\; =
\begin{bmatrix}
a_{11} & \cdots & a_{1n} \\
\vdots & \ddots & \vdots \\
a_{n1} & \cdots & a_{nn}
\end{bmatrix}
\odot
\begin{bmatrix}
x_1 x_1^* & \cdots & x_n x_1^* \\
\vdots & \ddots & \vdots \\
x_1 x_n^* & \cdots & x_n x_n^*
\end{bmatrix} \\
& \quad \quad \quad \quad \;\;\; =
\begin{bmatrix}
a_{11} x_1 x_1^* & \cdots & a_{1n} x_n x_1^* \\
\vdots & \ddots & \vdots \\
a_{n1} x_1 x_n^* & \cdots & a_{nn} x_n x_n^*
\end{bmatrix}
\end{aligned}
\end{equation}
Finally, the values resulting from equation (\ref{B3}) are equal to those from equation (\ref{B2}), and thus confirming that equation (\ref{B1}) holds.

\subsection{Evaluate the $\tilde{S}_1(\mathbf{Q}^{(i)}, \mathbf{W}^{(i)})$}\label{Appendix3}
In order to solve the optimization problem, it is necessary to compute the derivatives of matrices w.r.t. the matrix variables in the problem, and these calculations are examined in this appendix. First, according to reference \cite{ref34}, the general formulas related to the element-wise product are presented in equation (\ref{C1}):
\begin{equation}\label{C1}
\begin{aligned}
1) & \quad \mathbf{A} \odot \mathbf{B} = \mathbf{B} \odot \mathbf{A} , \quad \mathbf{A}, \mathbf{B} \in \mathds{C}^{m \times m} \\
2) & \quad (\mathbf{A} \odot \mathbf{B})^H = \mathbf{A}^H \odot \mathbf{B}^H = \mathbf{B}^H \odot \mathbf{A}^H = (\mathbf{B} \odot \mathbf{A})^H , \quad \mathbf{A}, \mathbf{B} \in \mathds{C}^{m \times m} \\
3) & \quad (\mathbf{A} \odot \mathbf{B})(\mathbf{C} \odot \mathbf{D}) = (\mathbf{A}\mathbf{C}) \odot (\mathbf{B}\mathbf{D}) , \quad \mathbf{A}, \mathbf{B}, \mathbf{C}, \mathbf{D} \in \mathds{C}^{m \times m}
\end{aligned}
\end{equation}
Now, according to references \cite{ref35, ref36}, the required derivatives in this study are generally computed as shown below,
\begin{equation}\label{C2}
\begin{aligned}
1) & \quad \frac{\partial}{\partial \mathbf{X}} \text{Tr}(\mathbf{A} \odot \mathbf{X}) = \text{diag}^* (\mathbf{A}), \\
2) & \quad \frac{\partial}{\partial \mathbf{X}} \text{Tr}(\mathbf{A} \odot \mathbf{X}^H) = \text{diag}^* (\mathbf{A})^H, \\
3) & \quad \frac{\partial}{\partial \mathbf{X}} \text{Tr}(\mathbf{A} \odot \mathbf{X}\mathbf{X}^H) = 2 \text{diag}^* (\mathbf{A}) \mathbf{X}, \\
4) & \quad \frac{\partial}{\partial \mathbf{X}} \text{Tr}(\mathbf{A}\mathbf{X}\mathbf{B}) = \mathbf{A}^T \mathbf{B}^T, \\
5) & \quad \frac{\partial}{\partial \mathbf{X}} \text{Tr}(\mathbf{A}\mathbf{X}^T\mathbf{B}) = \mathbf{B}\mathbf{A}, \\
6) & \quad \frac{\partial}{\partial \mathbf{X}} \text{Tr}(\mathbf{A}\mathbf{X}\mathbf{B}\mathbf{X}^T\mathbf{C}) = \mathbf{A}^T \mathbf{C}^T \mathbf{X}\mathbf{B}^T + \mathbf{CAXB}
\end{aligned}
\end{equation}
where the symbol $\text{diag}^*(\mathbf{A})$ means setting the off-diagonal elements of matrix $\mathbf{A}$ to zero. Now, based on equation (\ref{C2}), the final goal is to compute the value of $\tilde{S}_1(\mathbf{Q}^{(i)}, \mathbf{W}^{(i)})$, which is shown in equation (\ref{C3}):
\begin{equation}\label{C3}
\begin{aligned}
\tilde{S}_1(\mathbf{Q}, \mathbf{W}) &= \frac{1}{2} \| \tilde{\mathbf{H}}_I \odot \mathbf{Q} + \mathbf{H}_B \mathbf{W} \mathbf{H}_B^H \|_F^2 \\
\;\; &\geq \frac{1}{2} \| \tilde{\mathbf{H}}_I \odot \mathbf{Q}^{(i)} + \mathbf{H}_B \mathbf{W}^{(i)} \mathbf{H}_B^H \|_F^2 \\
&\quad + \text{Tr}(\nabla_{\mathbf{Q}}^H (\frac{1}{2} \| \tilde{\mathbf{H}}_I \odot \mathbf{Q}^{(i)} + \mathbf{H}_B \mathbf{W}^{(i)} \mathbf{H}_B^H \|_F^2)) (\mathbf{Q} - \mathbf{Q}^{(i)}) \\
&\quad + \text{Tr}(\nabla_{\mathbf{W}}^H (\frac{1}{2} \| \tilde{\mathbf{H}}_I \odot \mathbf{Q}^{(i)} + \mathbf{H}_B \mathbf{W}^{(i)} \mathbf{H}_B^H \|_F^2)) (\mathbf{W} - \mathbf{W}^{(i)}) \\
&\equiv \tilde{S}_1(\mathbf{Q}^{(i)}, \mathbf{W}^{(i)})
\end{aligned}
\end{equation}
To calculate $\tilde{S}_1(\mathbf{Q}^{(i)}, \mathbf{W}^{(i)})$, auxiliary matrices $\mathcal{A}$ and $\mathcal{B}$ are first defined and computed, and the value of $\tilde{S}_1(\mathbf{Q}^{(i)}, \mathbf{W}^{(i)})$ will be finally calculated. Below, the definition of the auxiliary matrix $\mathcal{A}$ is provided:
\begin{equation}\label{C4}
\begin{aligned}
\mathcal{A} &= \nabla_{\mathbf{Q}}^H (\frac{1}{2} \| \tilde{\mathbf{H}}_I \odot \mathbf{Q}^{(i)} + \mathbf{H}_B \mathbf{W}^{(i)} \mathbf{H}_B^H \|_F^2) \\
&= \nabla_{\mathbf{Q}}^H \left( \frac{1}{2} \text{Tr}((\tilde{\mathbf{H}}_I \odot \mathbf{Q}^{(i)} + \mathbf{H}_B \mathbf{W}^{(i)} \mathbf{H}_B^H) (\tilde{\mathbf{H}}_I \odot \mathbf{Q}^{(i)} + \mathbf{H}_B \mathbf{W}^{(i)} \mathbf{H}_B^H)^H) \right)
\end{aligned}
\end{equation}
Next, for simplicity in the calculations, the auxiliary variable $\mathbf{B}$ is defined as $\mathbf{B} = \mathbf{H}_B \mathbf{W}^{(i)} \mathbf{H}_B^H$. Now,
\begin{equation}\label{C5}
\begin{aligned}
\mathcal{A} &= \nabla_{\mathbf{Q}}^H \left( \frac{1}{2} \text{Tr}((\tilde{\mathbf{H}}_I \odot \mathbf{Q}^{(i)}) (\tilde{\mathbf{H}}_I \odot \mathbf{Q}^{(i)})^H + (\tilde{\mathbf{H}}_I \odot \mathbf{Q}^{(i)}) \mathbf{B}^H + \mathbf{B} (\tilde{\mathbf{H}}_I \odot \mathbf{Q}^{(i)})^H) \right) \\
&= \nabla_{\mathbf{Q}}^H \left( \frac{1}{2} \text{Tr}((\tilde{\mathbf{H}}_I \tilde{\mathbf{H}}_I^H \odot \mathbf{Q}^{(i)} (\mathbf{Q}^{(i)})^H) + (\tilde{\mathbf{H}}_I \odot \mathbf{Q}^{(i)}) \mathbf{B}^H + \mathbf{B} (\tilde{\mathbf{H}}_I \odot \mathbf{Q}^{(i)})^H) \right)
\end{aligned}
\end{equation}
Also, two other variables are defined as follows: $\mathbf{J}_R \in \mathds{C}^{2R \times 2R}$, where all elements of this matrix are equal to 1, and $\mathbf{H} = \tilde{\mathbf{H}}_I \tilde{\mathbf{H}}_I^H \in \mathds{C}^{2R \times 2R}$. Then,
\begin{equation}\label{C6}
\begin{aligned}
\mathcal{A} &= \nabla_{\mathbf{Q}}^H \left( \frac{1}{2} \text{Tr}((\mathbf{H} \odot \mathbf{Q}^{(i)} (\mathbf{Q}^{(i)})^H) + (\tilde{\mathbf{H}}_I \mathbf{B}^H \odot \mathbf{Q}^{(i)}) + (\mathbf{B} \odot \mathbf{J}_R) (\tilde{\mathbf{H}}_I \odot \mathbf{Q}^{(i)})^H) \right) \\
&= \frac{1}{2} ( 2 \text{diag}^* (\mathbf{H}) \mathbf{Q}^{(i)} + \text{diag}^* (\tilde{\mathbf{H}}_I \mathbf{B}^H) + \text{diag}^* (\mathbf{B} \tilde{\mathbf{H}}_I^H)^H )^H \\
&= (\text{diag}^* (\tilde{\mathbf{H}}_I \tilde{\mathbf{H}}_I^H) \mathbf{Q}^{(i)} + \text{diag}^* (\tilde{\mathbf{H}}_I \mathbf{B}^H))^H \\
&= (\mathbf{Q}^{(i)})^H \text{diag}^* (\tilde{\mathbf{H}}_I \tilde{\mathbf{H}}_I^H) + \text{diag}^* (\mathbf{B} \tilde{\mathbf{H}}_I^H)
\end{aligned}
\end{equation}
Finally, the auxiliary matrix $\mathcal{A}$ is computed and shown below:
\begin{equation}\label{C7}
\mathcal{A} = (\mathbf{Q}^{(i)})^H \text{diag}^* (\tilde{\mathbf{H}}_I \tilde{\mathbf{H}}_I^H) + \text{diag}^* (\mathbf{H}_B \mathbf{W}^{(i)} \mathbf{H}_B^H \tilde{\mathbf{H}}_I^H)
\end{equation}
Now, the value of the auxiliary matrix $\mathcal{B}$ should be computed as below,
\begin{equation}\label{C8}
\mathcal{B} = \nabla_{\mathbf{W}}^H \left( \frac{1}{2} \| \tilde{\mathbf{H}}_I \odot \mathbf{Q}^{(i)} + \mathbf{H}_B \mathbf{W} \mathbf{H}_B^H \|_F^2 \right)
\end{equation}
Next, the auxiliary variable $\mathbf{A}$ is defined as $\mathbf{A} = \tilde{\mathbf{H}}_I \odot \mathbf{Q}^{(i)} \in \mathds{C}^{2R \times 2R}$ to simplify the derivative calculations. So,
\begin{equation}\label{C9}
\begin{aligned}
&\mathcal{B} = \nabla_{\mathbf{W}}^H \left( \frac{1}{2} \text{Tr}\left((\mathbf{A} + \mathbf{H}_B \mathbf{W}^{(i)} \mathbf{H}_B^H) (\mathbf{A} + \mathbf{H}_B \mathbf{W}^{(i)} \mathbf{H}_B^H)^H\right) \right)\\
          &\;\;= \nabla_{\mathbf{W}}^H \left( \frac{1}{2} \text{Tr}\left(\mathbf{A} \mathbf{H}_B \mathbf{W}^{(i)} \mathbf{H}_B^H + \mathbf{A}^H \mathbf{H}_B \mathbf{W}^{(i)} \mathbf{H}_B^H + \mathbf{H}_B \mathbf{W}^{(i)} \mathbf{H}_B^H \mathbf{H}_B \mathbf{W}^{(i)} \mathbf{H}_B^H\right) \right)\\
          &\;\;= \left( \mathbf{H}_B^H \mathbf{A} \mathbf{H}_B + \mathbf{H}_B^H \mathbf{H}_B \mathbf{W}^{(i)} \mathbf{H}_B^H \mathbf{H}_B \right)^H
\end{aligned}
\end{equation}
Finally, the auxiliary matrix $\mathcal{B}$ is computed and shown below:
\begin{equation}\label{C10}
\mathcal{B} = \left( \mathbf{H}_B^H (\tilde{\mathbf{H}}_I \odot \mathbf{Q}^{(i)})^H \mathbf{H}_B + \mathbf{H}_B^H \mathbf{H}_B \mathbf{W}^{(i)} \mathbf{H}_B^H \mathbf{H}_B \right)
\end{equation}
In conclusion, the final expression for $\tilde{S}_1(\mathbf{Q}^{(i)}, \mathbf{W}^{(i)})$ based on the final values of the auxiliary matrices $\mathcal{A}$ and $\mathcal{B}$ is computed as shown in equation (\ref{C11}):
\begin{equation}\label{C11}
\begin{aligned}
\tilde{S}_1(\mathbf{Q}^{(i)}, \mathbf{W}^{(i)}) &= \frac{1}{2} \| \tilde{\mathbf{H}}_I \odot \mathbf{Q}^{(i)} + \mathbf{H}_B \mathbf{W}^{(i)} \mathbf{H}_B^H \|_F^2 \\
& \quad + \text{Tr}(((\mathbf{Q}^{(i)})^H \text{diag}^* (\tilde{\mathbf{H}}_I \tilde{\mathbf{H}}_I^H) + \text{diag}^* (\mathbf{H}_B \mathbf{W}^{(i)} \mathbf{H}_B^H \tilde{\mathbf{H}}_I^H))(\mathbf{Q} - \mathbf{Q}^{(i)})) \\
& \quad + \text{Tr}((\mathbf{H}_B^H (\tilde{\mathbf{H}}_I \odot \mathbf{Q}^{(i)})^H \mathbf{H}_B + \mathbf{H}_B^H \mathbf{H}_B \mathbf{W}^{(i)} \mathbf{H}_B^H \mathbf{H}_B)(\mathbf{W} - \mathbf{W}^{(i)}))
\end{aligned}
\end{equation}

\end{document}